\documentclass[11pt,a4paper]{article}
\usepackage[utf8x]{inputenc}
\usepackage[T1]{fontenc}
\usepackage{tikz}
\usepackage{tikz-3dplot}
\usetikzlibrary{arrows.meta,calc,intersections,math}
\tdplotsetmaincoords{60}{30}

\usepackage{cite}
\usepackage{hyperref}

\usepackage[margin=25truemm]{geometry}
\usepackage{graphicx}
\usepackage{calc} 
\usepackage{authblk}
\usepackage{bm}
\usepackage{physics}
\usepackage{enumitem} 
\usepackage{braket,mathtools,amsmath,amssymb,subcaption}
\usepackage{mathrsfs}
\usepackage{authblk}
\usepackage[all]{nowidow} 
\usepackage{url}

\usepackage{lipsum} 

\usepackage{framed} 
\usepackage[framed]{ntheorem}
\newframedtheorem{frm-thm}{Theorem}

\def\m{\mathcal}

\def\d{\delta}

\def\s{\sigma}
\def\o{\omega}

\def\hx{\hat{x}}
\def\hy{\hat{y}}
\def\hz{\hat{z}}
\def\exx{\mathbf{e}_x}
\def\eyy{\mathbf{e}_y}

\def\br{\mathbf{r}}

\def\bx{\mathbf{l}_x}
\def\by{\mathbf{l}_y}
\def\bz{\mathbf{l}_z}
\def\bl{\mathbf{l}}
\def\bc{\mathbf{c}}
\def\bp{\mathbf{p}}

\def\ex{\frac{\mathbf{e}_x}{2}}
\def\ey{\frac{\mathbf{e}_y}{2}}
\def\ez{\frac{\mathbf{e}_z}{2}}

\usepackage{hyperref}
\hypersetup{colorlinks,linkcolor=cyan,citecolor=blue}

\usepackage{tikz}  
\usetikzlibrary{arrows.meta,positioning,calc}
\begin{document} 

\title{Non-invertible translation from Lieb–Schultz–Mattis anomaly }
\author{Tsubasa Oishi$^1$}
\author{Takuma Saito$^1$}
\author{Hiromi Ebisu$^2$}

\affil{$^1$Yukawa Institute for Theoretical Physics, Kyoto University, Kyoto 606-8502, Japan}
  \affil{$^2$Interdisciplinary Theoretical and Mathematical Sciences Program (iTHEMS)~RIKEN, Wako 351-0198, Japan}  
  
\maketitle
\thispagestyle{empty}

\begin{abstract}
Symmetry provides powerful non-perturbative constraints in quantum many-body systems. A prominent example is the Lieb–Schultz–Mattis (LSM) anomaly—~a mixed ’t Hooft anomaly between internal and translational symmetries that forbids a trivial symmetric gapped phase. In this work, we investigate lattice translation operators in systems with an LSM anomaly. We construct explicit lattice models in two and three spatial dimensions and show that, after gauging the full internal symmetry, translation becomes non-invertible and fuses into defects of the internal symmetry. 
The result is supported by
the anomaly-inflow in view of topological field theory.
Our work extends earlier one-dimensional observations to a unified higher-dimensional framework and clarifies their origin in mixed anomalies and higher-group structures, highlighting a coherent interplay between internal and crystalline symmetries.
\end{abstract}

\newpage
\pagenumbering{arabic}

\tableofcontents
\section{Introduction}
Symmetry is one of the foundational organizing principles of physics, providing non-perturbative constraints on quantum many-body systems and enabling deeper insights across condensed matter, high-energy physics, and quantum information. A central tool in this context is the notion of an ’t Hooft anomaly, which characterizes obstructions to gauging global symmetries and 
underlies the modern classification of symmetry-protected topological (SPT) phases~\cite{spt2013,1986LMaPh..12...57A,PhysRevB.86.115109,Kapustin2014}.

Among crystalline symmetries, lattice translation occupies a distinguished position. It imposes strong constraints on the spectrum and possible phases of matter. A paradigmatic example is the Lieb–Schultz–Mattis~(LSM) theorem~\cite{Lieb1961,PhysRevLett.84.1535,PhysRevB.69.104431,Cheng:2015kce}, originally formulated for spin-1/2 chains, which states that a system with both internal and translational symmetries cannot support a trivial symmetric gapped ground state. Recent developments have reinterpreted the LSM theorem as a mixed ’t Hooft anomaly between the internal symmetry and the lattice translation symmetry—now commonly referred to as the \textit{LSM anomaly}. This anomaly-based viewpoint has triggered substantive conceptual advances. In particular, the LSM anomaly has been shown to organize higher-group symmetry structures relating internal and crystalline symmetries, and to naturally give rise to the notion of \textit{modulated symmetry}, where the action of the internal symmetry is “twisted’’ by lattice translation~\cite{2024multipole,Pace:2024acq,Pace:2025hpb,bulmash2025defect,Ebisu:2025mtb}. 
Typical examples of such symmetries are dipole symmetries, corresponding to conservation of dipole moments~\cite{griffin2015scalar,Pretko:2018jbi,PhysRevX.10.011047,Gorantla:2022eem}. 
These ideas have become increasingly relevant across multiple communities, including quantum information, topological field theory, and the study of exotic quantum phases~\cite{Haah2011,chamon,Vijay}.
 \par

It has been clarified that gauging a \emph{subgroup} of the internal symmetry gives rise to modulated symmetries~\cite{Aksoy:2023hve,Seifnashri:2023dpa,Pace:2025hpb,Ebisu:2025mtb}. However, physics after gauging \emph{full} internal symmetry has not been explored in a systematic manner.
In this paper, we 
address this question and show that gauging full internal symmetry inevitably promotes lattice translation itself to a non-invertible symmetry operator~(See also Fig.~\ref{diagram}.). Our results imply that, in the presence of an LSM anomaly, lattice translation cannot be consistently realized as an invertible symmetry once all internal symmetries are fully gauged. 
We construct explicit two- and three-dimensional lattice models exhibiting the LSM anomaly and demonstrate that lattice translation becomes \textit{non-invertible} after gauging the full internal symmetry. 
These non-invertible translations fuse into internal symmetry defects (more precisely, defects of the dual internal symmetries). The resulting fusion rules, defect algebra, and anomaly inflow are further validated by a complementary continuum field~theoretical analysis. While non-invertible translations were previously noticed in one dimensional chains~\cite{Seifnashri:2023dpa, Pace:2024acq}, our work gives the first unified and higher-dimensional framework, clarifying their origin from mixed anomalies and higher group structures. 

We further point out that the resulting non-invertible translation operators admit a natural interpretation in terms of $N$-ality defects. In particular, when the underlying LSM anomaly corresponds to the so-called “type-III anomaly’’~\cite{deWildPropitius:1995cf,wang2015field}, the non-invertible translation operator realizes an $N$-ality defect whose fusion rules and grading are fixed by the anomaly structure\footnote{The $N$-ality symmetries are studied in various contexts. See, e.g.,~Refs~\cite{Lu:2022ver,Lu:2024lzf, Lu:2024ytl, Ando:2024hun, Maeda:2025rxc, Kaidi:2025hyr, Lu:2025gpt, Thorngren:2021yso, Choi:2022jqy, Choi:2022zal, Cordova:2022ieu, Hayashi:2022fkw}.}. From this viewpoint, the non-invertibility of translation is not an accidental lattice artifact, rather, a direct consequence of anomaly inflow that necessitates additional topological degrees of freedom (d.o.f) living on the translational defect. Altogether, our results establish a unified framework in which modulated symmetries and non-invertible translations arise as different facets of LSM anomalies. By explicitly constructing lattice models and complementary anomaly inflow descriptions, we elucidate how internal and crystalline symmetries intertwine to produce intrinsically non-invertible symmetry operators in anomalous quantum phases.
\par

The rest of this work is organized as follows. In Sec.~\ref{sec2}, we go over the non-invertible translation operators in a one dimensional spin chain studied in Refs.~\cite{Seifnashri:2023dpa, Pace:2024acq}. In Sec.~\ref{sec3}, we introduce a two dimensional lattice spin systems with the LSM anomaly and study the non-invertible translation operators. Sec.~\ref{sec4} is devoted to complementary field theories describing the non-invertible translation operators. Finally, in Sec.~\ref{sec5}, we conclude our works with a few remarks. Technical details are relegated to appendices.

\section{Review of non-invertible translation symmetries in 1D}\label{sec2}
In this section, we review the non-invertible translation symmetries obtained by gauging the full internal symmetries of a 1D system with the LSM anomaly~\cite{Seifnashri:2023dpa, Pace:2024acq}. We mainly follow argument presented in~\cite{Pace:2024acq}. Furthermore, we construct the non-invertible translation operators and compute their fusion rules. As far as we are aware, while the non-invertible translation was noticed in several works~\cite{Seifnashri:2023dpa, Pace:2024acq}, a thorough discussion on this operator, including details of the fusion rules, is not discussed previously. 

We start by defining the $\mathbb{Z}_N$-based XZ model, whose Hamiltonian on a periodic chain with system size $L$ is given by 
\begin{equation}
    H_{\mathrm{XZ}} = \sum_{j=1}^L (J_\mathrm{X}X_jX_{j+1}^{\dagger} + J_\mathrm{Z}Z_jZ_{j+1}^{\dagger}) + h.c., \label{XZ model}
\end{equation}
where $X_j$ and $Z_j$ are $\mathbb{Z}_N$ shift and clock operators at site $j$, which satisfy the relation
\begin{equation}
    Z_iX_j=\omega^{\delta_{ij}} X_jZ_i, \quad \omega=e^{\frac{2\pi i}{N}}, \quad X_j^N=Z_j^N=1.
\end{equation}
This model has two $\mathbb{Z}_N$ $0$-form symmetries generated by
\begin{equation}
    U_X=\prod_{j=1}^LX_j, \quad U_Z=\prod_{j=1}^LZ_j, \label{two 0-form}
\end{equation}
and lattice translation symmetry given by\footnote{This product $\prod_j$ is defined as the ordered product, i.e. $\prod_{j=1}^{L-1}a_j=a_1a_2\cdots a_{L-1}$.}
\begin{equation}
    T_x=\prod_{j=1}^{L-1}T_j, \quad T_j=\frac{1}{N}\sum_{a,b=0}^{N-1}\omega^{ab}(X_jX_{j+1}^{\dagger})^a(Z_jZ_{j+1}^{\dagger})^b \label{translation}
\end{equation}
whose action on a local operator reads
$T_x X_{j}=X_{j+1}T_x$ and similarly for $Z_j$. 
Further, this model is a typical example with an LSM anomaly, and has the following projective algebra
\begin{equation}
    U_ZU_X=\omega^LU_XU_Z,
\end{equation}
with an anomalous phase $\omega^L$ depending on the system size $L$. This projective algebra characterizes an LSM anomaly between two internal $\mathbb{Z}_N$ symmetries and lattice translation symmetry \cite{Seifnashri:2023dpa, Aksoy:2023hve, Alavirad:2019iea}. The LSM anomaly is also manifested by the projective representation of $\mathbb{Z}_N\times\mathbb{Z}_N$ at each site. Throughout this section, we assume that the system size $L$ is a multiple of $N$. 

In what follows, we show that non-invertible translation symmetry emerges when gauging the full internal symmetries \eqref{two 0-form}. To see this, we perform gauging of the full symmetries~\eqref{two 0-form} each by each consisting of the two steps.
In the first step, we gauge one $Z_N$ $0$-form symmetry generated by $U_Z$. The gauged model respects a dipole symmetry~\cite{Seifnashri:2023dpa, Aksoy:2023hve, Cao:2024qjj}, which is discussed in Sec.~\refeq{Sec2.1}. We then move on to the second step, which is gauging the dipole symmetry obtained in the first step to see how non-invertible translation has emerged~(Sec.~\refeq{Sec2.2}). This two-step procedure amounts to simultaneously gauging two $\mathbb{Z}_N$ 0-form symmetries~\eqref{two 0-form}.

\subsection{Dipole symmetry from LSM anomaly} \label{Sec2.1}
In this subsection, we start with gauging one $\mathbb{Z}_N$ $0$-form symmetry generated by $U_Z$. To do this, we introduce extended Hilbert space on each link, and gauge variables $\s^X_{(j,j+1)}$ and $\s^Z_{(j,j+1)}$ acting on each link. These variables satisfy
\begin{equation*}  \s^Z_{(j,j+1)}\s^X_{(k,k+1)}=\o^{\d_{jk}}\s^X_{(k,k+1)}\s^Z_{(j,j+1)}. 
\end{equation*}
We define the Gauss's law operator as
\begin{equation}
    G_{j} = \s^{Z}_{(j-1,j)}Z_j\s^{Z\dagger}_{(j,j+1)}.\label{gauss}
\end{equation}
The Gauss's law operator generates a local $\mathbb{Z}_N$ symmetry, 
which can be understood by
\begin{equation}
    \prod_{j=1}^LG_j=U_Z, \quad G_j^N=1.
\end{equation}
Intuitively, we decompose the global symmetry into local ones. This is consistent with notion of gauging where a global symmetry is promoted to local one. \par
The gauge-invariant Hamiltonian is obtained by minimally coupling the system to the gauge variables so that it commutes with the Gauss's law~\eqref{gauss}. Under this procedure, the gauged Hamiltonian \eqref{XZ model} reads 
\begin{equation}
    H_{\mathrm{dipole}}:=H_{\mathrm{XZ}/\mathbb{Z}_N} = \sum_{j=1}^L (J_\mathrm{X}X_j\s^X_{(j,j+1)}X_{j+1}^{\dagger} + J_\mathrm{Z}Z_jZ_{j+1}^{\dagger}) + h.c.,\label{dipole}
\end{equation}
which is gauge invariant since it commutes with the Gauss's law operator. Finally, we must impose the Gauss law constraint that the physical states $\ket{\psi}$ should be invariant under the action of $G_j$, namely $G_j\ket{\psi}=\ket{\psi}, \forall j$. Furthermore, to solve the Gauss's law constraint, we implement the unitary transformation on local operators so that
\begin{equation}
    Z_j \rightarrow \s^{Z\dagger}_{(j-1,j)}Z_j\s^{Z}_{(j,j+1)}, \quad 
    \sigma^X_{(j,j+1)} \rightarrow X_j^\dagger\sigma^X_{(j,j+1)}X_{j+1}.
\end{equation}
Under this transformation, the Gauss's law operator $G_j$ is transformed into $Z_j$, and thus the Gauss's law constraint becomes $Z_j=1~\forall j$.
Thus, the gauged Hamiltonian~\eqref{dipole} is equivalent to 
\begin{equation}
    H_{\mathrm{dipole}} = \sum_{j=1}^L \big(J_\mathrm{X}\s^X_{(j,j+1)} + J_\mathrm{Z}\s^{Z\dagger}_{(j-1,j)}(\s^{Z}_{(j,j+1)})^2\s^{Z\dagger}_{(j+1,j+2)}\big) + h.c.. \label{dipole Hamiltonian}
\end{equation}
This Hamiltonian has a dipole symmetry generated by 
\begin{equation}
    Q_0=\prod_{j=1}^L\s^X_{(j,j+1)}, \quad Q_x=\prod_{j=1}^L\big(\s^X_{(j,j+1)}\big)^j,
\end{equation}
where $Q_0$ generates the dual $\mathbb{Z}_N$ 0-form symmetry and $Q_x$ generates the $\mathbb{Z}_N$ 0-form dipole symmetry. These symmetries, including the lattice translation operator $T_x$, form the dipole algebra
\begin{equation}
    T_xQ_xT_x^{-1}=Q_0^\dagger Q_x,\quad T_xQ_0T_x^{-1}=Q_0, \label{dipole0}
\end{equation}
where $T_x$ acts on a local operator $O_j$ by shifting one lattice site, i.e.,  $T_xO_jT_x^{-1}=O_{j+1}$. Note that since the translation operator \eqref{translation}  is gauge invariant\footnote{Here, the gauge invariance of the translation symmetry means that $T_xG_jT_x^{-1}=G_{j+1}$}, it remains a symmetry. The relation~\eqref{dipole0} exhibits a hierarchical structure of dipole and global charges related via a translation operator that we dub \emph{dipole algebra}.

The emergence of the dipole symmetry after gauging can be understood as follows. Before gauging, the operator $U_X$ can be rewritten as $\prod_{j=1}^{L-1}\big(X_jX_{j+1}^\dagger\big)^j$. By minimally coupling and performing the unitary transformation, we find that $U_X$ is mapped to the dipole symmetry~$Q_x$ in the gauged theory. It is worth addressing that generally when the system has LSM anomaly with two internal global and translational ones, and gauge one of the internal symmetries, the modulated symmetries have emerged.


\subsection{Non-invertible translation symmetry from dipole symmetry} \label{Sec2.2}
In this subsection, we gauge the dipole symmetry generated by $Q_x$. General procedure for gauging of other types of modulated symmetries on lattice has been studied in e.g, Refs~\cite{Gorantla:2022eem,Pace:2024acq,Ebisu_BO_2025}. 
For simplicity, we focus on the case $N=2$. The analysis for prime $N=p$ is relegated to the Appendix.~\ref{appendix.A}.

As in the previous subsection, we introduce extended Hilbert space on each vertex, and gauge variables $\m{X}_j$ and $\m{Z}_j$ acting on each vertex. 
We define the Gauss's law operator as
\begin{equation}
    G_{(j.j+1)}=\m{X}_j\big(\s^X_{(j,j+1)}\big)^j\m{X}_{j+1}, 
\end{equation}
and impose the Gauss's law constraint, namely, $G_{(j.j+1)}=1$ at each link. 
By minimally coupling to \eqref{dipole Hamiltonian}, we obtain the gauge invariant Hamiltonian
\begin{equation}
    H_{\mathrm{XZ}/\mathbb{Z}_2\times\mathbb{Z}_2} = \sum_{j=1}^L \big(J_\mathrm{X}\s^X_{(j,j+1)} + J_\mathrm{Z}\s^{Z}_{(j-1,j)}\m{Z}_j^{-j+1}\m{Z}_{j+1}^{j+1}\s^{Z}_{(j+1,j+2)}\big) + h.c.. \label{N=2 gauged dipole Hamiltonian}
\end{equation}
One can readily find that this model has two $\mathbb{Z}_2$ $0$-form symmetries generated by
\begin{equation}
    \widehat{U}_X=\prod_{j=1}^{L}\m{Z}_j, \quad \widehat{U}_Z=\prod_{j=1}^{L}\s^X_{(j,j+1)}, \label{Z_2 sym}
\end{equation}
where these labels in $\widehat{U}_X$ and $\widehat{U}_Z$ indicate that they originate from the original $U_X$ and $U_Z$. However, this Hamiltonian does not possess the lattice translation symmetry, since lattice translation operator is not gauge invariant\footnote{Note that $T_x^{2\alpha}(\alpha=1,2,\cdots, \ell-1)$ remain as ordinary lattice translation symmetry since they are gauge invariant, where the system size $L=2\ell$.}, 
\begin{equation}
    T_xG_{(j,j+1)}T_x^{-1}\ne G_{(j+1,j+2)}.
\end{equation}
To make translation operator gauge invariant, we modify the operator as
\begin{equation}
    T_x\rightarrow \mathcal{T}_x := T_xW,\label{TST}
\end{equation}
where $W$ implements the transformation\footnote{The explicit form of the operator is given in Ref \cite{Seifnashri:2024dsd}, and the generalization to 
a matrix product operator expression on an arbitrary lattice is given in Ref~\cite{Cao:2025qhg}. }
\begin{equation}
    \begin{aligned} 
        \s^Z_{(j,j+1)}\s^{Z}_{(j+1,j+2)} &\rightarrow \s^Z_{(j,j+1)}\m{Z}_{j+1}\s^{Z}_{(j+1,j+2)} \\
        \m{X}_j\m{X}_{j+1} &\rightarrow \m{X}_{j}\s^X_{(j,j+1)}\m{X}_{j+1}.
    \end{aligned}
\end{equation}
This transformation can be interpreted as the topological manipulation TST, where S and T represent the gauging of two $\mathbb{Z}_2$ 0-form symmetries \eqref{Z_2 sym} and the stacking of SPT phase protected by the same symmetry, respectively. This procedure is referred to as the Kennedy-Tasaki transformation \cite{Kennedy:1992ifl, Kennedy:1992tke, Oshikawa:1992smv, Li:2023ani}. Since $W$ involves a gauging operation, it is non-invertible. Indeed, it is known that the operator realizing TST obeys the non-invertible fusion rule given by
\begin{eqnarray} \label{fusion of W}
    W \times W = (1 + \widehat{U}_X)(1 + \widehat{U}_Z),
\end{eqnarray}
see, e.g., Ref.\cite{Seifnashri:2024dsd}.


The modified translation operator \eqref{TST} is now gauge invariant and commutes with the Hamiltonian. However, it 
becomes \emph{non-invertible}
due to $W$. Indeed, using \eqref{fusion of W}, we find that
the fusion rule of modified translation operator is given by
\begin{equation} \label{fusion}
    \mathcal{T}_x \times \mathcal{T}_x^{\dagger} = (1 + \widehat{U}_X)(1 + \widehat{U}_Z),
\end{equation}
where $\mathcal{T}_x^{\dagger}:=WT_x^{-1}$ is the modified translation operator in the negative direction. This fusion rule indicates that 
translation operator is non-invertible.

\section{Two dimensional lattice model}\label{sec3}
\begin{figure}
    \begin{center}
         \begin{subfigure}[h]{0.59\textwidth}
       \centering
  \includegraphics[width=0.9\textwidth]{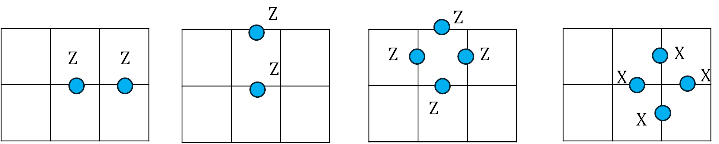}
         \caption{}\label{lsm1}
             \end{subfigure}
            \begin{subfigure}[h]{0.40\textwidth}
            \centering
  \includegraphics[width=1.0\textwidth]{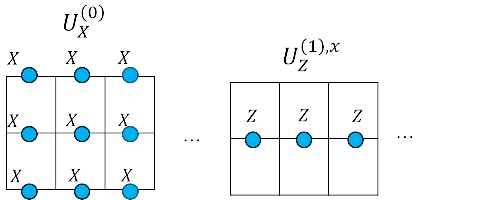}
         \caption{}\label{0formdual}
             \end{subfigure}
                        \begin{subfigure}[h]{0.68\textwidth}
            \centering
  \includegraphics[width=1.0\textwidth]{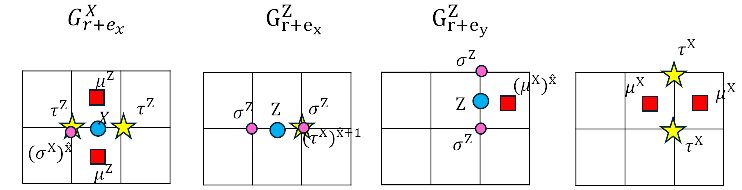}
         \caption{}\label{gs}
             \end{subfigure}
              \begin{subfigure}[h]{0.78\textwidth}
                         \centering
  \includegraphics[width=1.3\textwidth]{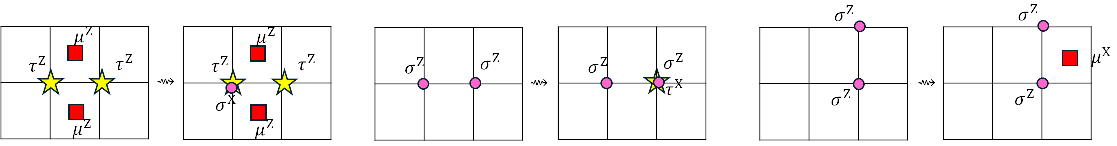}
         \caption{}\label{tst}
             \end{subfigure}
 \end{center}
 \caption{(a)~Four terms that constitute the Hamiltonian~\eqref{hamiltonian1}. (b) $0$-form and $1$-form symmetries that the model respects corresponding to~\eqref{symmetry}. (c)~Gauss's laws for $0$-form and $1$-form symmetries [first three terms, corresponding to~\eqref{g1}-\eqref{g3}] and gauge flux operator (last term). (d) Some examples of how operator $W$ (the TST transformation) defined in~\eqref{51}
 acts on operators. }
 \end{figure}

After reviewing non-invertible translation in one dimension, now we turn to main parts of this paper. In this section, we introduce a lattice model with the LSM anomaly in 2D and study lattice translational operators after gauging full internal symmetries. Throughout this section, we focus on the case $N=2$, that is, we deal with $\mathbb{Z}_2$ spins. 
We envisage square lattice where we have $\mathbb{Z}_2$ spin on each link. 
Denoting Pauli operator on each link as $X_{\bullet}$ and $Z_{\bullet}$, 
the Hamiltonian is defined by (see also Fig.~\ref{lsm1})
\begin{eqnarray}
    H_{2D}=&-&J_x\sum_{\br}Z_{\br-\ex}Z_{\br+\ex}-J_y\sum_{\br}Z_{\bp-\ey}Z_{\bp+\ey}\nonumber\\
      &-&J_G\sum_{\br}X_{\br-\ex}X_{\br+\ex}X_{\br-\ey}X_{\br+\ey}-J_B\sum_\bp Z_{\bp-\ex}Z_{\bp+\ex}Z_{\bp+\ey}Z_{\bp-\ey}.\label{hamiltonian1}
\end{eqnarray}
Here, we have introduced vectors, $\br\vcentcolon=(\hx,\hy)$, $\exx\vcentcolon=(1,0)$, $\eyy\vcentcolon=(0,1)$, and $\bp\vcentcolon=(\hx+\frac{1}{2},\hy+\frac{1}{2})$.
Also,~$\hx,\hy$ denote the coordinates of a node of the lattice, which take integer numbers. 
We impose the periodic boundary condition with the system size $L_x\times L_y$. 
The first two terms in~\eqref{hamiltonian1} describe standard nearest neighboring spin couplings in the horizontal and vertical directions, whereas the last two constitute the $\mathbb{Z}_2$ toric code~\cite{KITAEV20032}. In what follows, we take a limit $J_B\to\infty$ so that a fluxless condition is imposed, namely, 
we focus on a sector satisfying 
\begin{eqnarray}
    Z_{\bp-\ex}Z_{\bp+\ex}Z_{\bp+\ey}Z_{\bp-\ey}\ket{\Omega}=\ket{\Omega}\;\forall\bp.\label{flux}
\end{eqnarray}
 The model~\eqref{hamiltonian1} respects two global symmetries, $0$-form and $1$-form symmetries, described by (Fig.~\ref{0formdual})
 \begin{eqnarray}
     U^{(0)X}=\prod_{\br}X_{\br+\ex},\quad U^{(1)Z}_x=\prod_{\hx=1}^{L_x}Z_{(\hx+\frac{1}{2},1)},\quad U^{(1)Z}_y=\prod_{\hy=1}^{L_y}Z_{(1,\hy+\frac{1}{2})}.\label{symmetry}
 \end{eqnarray}
 While the $0$-form symmetry corresponds to spin flip on entire horizontal links, the $1$-form symmetry is described by noncontractible loop in $x$- or $y$-direction. 
 Furthermore, the model exhibits the LSM anomaly involving $0$-form, $1$-form, and translational symmetry in the $x$-direction: 
 \begin{eqnarray}
      U^{(0)X} U^{(1)Z}_x=(-1)^{L_x} U^{(1)Z}_x U^{(0)X}.\label{lsm}
 \end{eqnarray}

 When we gauge either $0$- or $1$-form symmetry, the $1$-form or $0$-form modulated symmetry (more precisely, dipole symmetry) in the $x$-direction~\cite{10.21468/SciPostPhys.14.5.106,Gorantla:2022pii} has emerged, respectively, which is consistent with the previous study~\cite{Ebisu:2025mtb}. For the sake of completeness, we accommodate Appendix.~\ref{appb} to elaborating on this point, to which interested readers can refer. For our purpose, we focus on gauging both of $0$- or $1$-form symmetries simultaneously to study how non-invertible translation symmetry appears. \par
 To proceed, we introduce extended Hilbert space defined on each node and plaquette whose Pauli operators are described by $\tau^{X/Z}_{\br}$, $\sigma^{X/Z}_{\br}$, $\mu^{X/Z}_{\bp}$
 impose the following Gauss's laws (Fig.~\ref{gs}):
 \begin{eqnarray}
1=\left(\sigma^X_{\br}\right)^{\hx}X_{\br+\ex}\tau^Z_{\br+\exx}\tau^Z_{\br}\mu^Z_{\br+\ex+\ey}\mu^Z_{\br+\ex-\ey}~\left( \vcentcolon=G^X_{\br+\ex}\right)\label{g1}\\
     1=Z_{\br+\ex}\sigma^Z_{\br}\sigma^Z_{\br+\exx}\left(\tau^X_{\br+\exx}\right)^{\hx+1}~\left(\vcentcolon= G^Z_{\br+\ex}\right)~\label{g2}\\
     1= ~Z_{\br+\ey}\sigma^Z_{\br}\sigma^Z_{\br+\eyy}\left(\mu^X_{\bp}\right)^{\hx+1}~\left(\vcentcolon=G^Z_{\br+\ey}\right)\label{g3}
 \end{eqnarray}
 Aside from the position dependent terms, namely, $\left(\sigma^X_{\br}\right)^{\hx}$, $\left(\tau^X_{\br+\exx}\right)^{\hx+1}$, and $\left(\mu^X_{\bp}\right)^{\hx+1}$, these three equations describe the standard Gauss's laws for $\mathbb{Z}_2$ $0$- and $1$-form symmetries. 
 The position dependent terms are introduced so that the three Gauss's laws~\eqref{g1}-\eqref{g3} commute with one another. 
Further, we impose the following condition:
\begin{eqnarray}
\prod_{\hx=1}^{L_x}\prod_{\hy=1}^{L_y}\left(\sigma^X_{(\hx,\hy)}\right)^{\hx}=1,\quad \prod_{\hx=1}^{L_x}\left(\tau^X_{(\hx+1,\hy)}\right)^{\hx+1}=1\quad\forall \hy,\quad  \prod_{\hy=1}^{L_y}\left(\mu^X_{(\hx+\frac{1}{2},\hy+\frac{1}{2})}\right)^{\hx}=1\quad\forall \hx,
\end{eqnarray}
ensuring that global symmetries~\eqref{symmetry} are decomposed into local ones, i.e., 
\begin{eqnarray*}
    \prod_{\hx=1}^{L_x}\prod_{\hy=1}^{L_y}G^X_{\br+\ex}=U^{(0)X},
\end{eqnarray*}
 and similarly for other symmetries. 
 \par
 To proceed, we modify the spin terms in the Hamiltonian so that they commute with the Gauss's laws~\eqref{g1}-\eqref{g3}, corresponding to the minimal coupling, viz
 \begin{eqnarray}
     Z_{\br-\ex} Z_{\br+\ex}\to   Z_{\br-\ex}\tau^X_{\br} Z_{\br+\ex},\quad Z_{\br+\ex} Z_{\br+\ex+\eyy}\to   Z_{\br+\ex}\mu^X_{\bp} Z_{\br+\ex+\eyy}\nonumber\\
     X_{\br-\ex}X_{\br+\ex}X_{\br-\ey}X_{\br+\ey}\to   X_{\br-\ex}X_{\br+\ex}X_{\br-\ey}X_{\br+\ey}\sigma^X_\br.\nonumber\\
      Z_{\bp-\ex}Z_{\bp+\ex}Z_{\bp+\ey}Z_{\bp-\ey}\to  Z_{\bp-\ex}Z_{\bp+\ex}Z_{\bp+\ey}Z_{\bp-\ey}\mu^X_{\bp}.
 \end{eqnarray}
We further add the following Gauge invariant flux 
\begin{eqnarray}
    -J_B^\prime \sum_{\br}\tau^X_{\br}\tau^X_{\br+\eyy}\mu^X_{\br-\ex+\ey}\mu^X_{\br+\ex+\ey}\label{flux2}
\end{eqnarray}
to the Hamiltonian~\eqref{hamiltonian1}
to ensure that the gauge theory does not admit excess flux.
From the Gauss's laws~\eqref{g2}~\eqref{g3}, the condition~\eqref{flux} is rewritten as 
\begin{eqnarray}
   Z_{\bp-\ex}Z_{\bp+\ex}Z_{\bp+\ey}Z_{\bp-\ey}\mu^X_{\bp}=1\quad\forall\bp\nonumber\\
   \Leftrightarrow \left(\tau^X_{\br}\tau^X_{\br+\eyy}\mu^X_{\br-\ex+\ey}\mu^X_{\br+\ex+\ey}\right)^{\hx+1}=1\quad\forall\br.
\end{eqnarray}
This condition is included in the term~\eqref{flux2} with $J_B^\prime\to\infty$. 
Hence, after gauging, the Hamiltonian~\eqref{hamiltonian1} becomes
\begin{eqnarray}
    {H}_{2D/\mathbb{Z}_2^{(1)}\times\mathbb{Z}_2^{(0)}}=&-&J_x\sum_{\br}Z_{\br-\ex}\tau^X_{\br} Z_{\br+\ex}-J_y\sum_{\br}Z_{\br+\ex}\mu^X_{\bp} Z_{\br+\ex+\eyy}\nonumber\\
      &-&J_G\sum_{\br} X_{\br-\ex}X_{\br+\ex}X_{\br-\ey}X_{\br+\ey}\sigma^X_\br  
      \nonumber\\
      &-&J_B^\prime \sum_{\br}\tau^X_{\br}\tau^X_{\br+\eyy}\mu^X_{\br-\ex+\ey}\mu^X_{\br+\ex+\ey}.\label{hamiltonian}
\end{eqnarray}
with $J^\prime_B\to\infty$ so that the gauged Hamiltonian is fluxless. 
The gauged Hamiltonian with Gauss's laws~\eqref{g1}-\eqref{g3} admits the following dual symmetries: 
\begin{eqnarray}
    \eta^{(0)}=\prod_{\hx=1}^{L_x}\prod_{\hy=1}^{L_y}\sigma^X_{\br},\quad \eta^{(1)}_x=\prod_{\hx=1}^{L_x}\tau^X_{(\hx,1)},\quad \eta^{(1)}_y=\prod_{\hy=1}^{L_y}\mu^X_{(\frac{1}{2},\hy+\frac{1}{2})}\label{loop}
\end{eqnarray}
which are $0$-and $1$-form symmetries, mirroring the fact that when gauging a $p$-form global symmetry in $(d+1)$ spacetime dimension, a dual $(d-p-1)$-form global symmetry has been emerged~\cite{Gaiotto:2014kfa}.
The last two terms in~\eqref{loop} are noncontractible loops of the gauge fields in the $x$- and $y$-direction. Note that these operators are topological, meaning, independent operators depend solely on the homology class in the ground state subspace, where there is no excess magnetic flux rather than on the full tensor-factorized Hilbert space. \par

We investigate fate of translation symmetry after gauging. To this end, 
we introduce a lattice translation operator in the $x$-direction, $T_x$ whose action on an operator as $T_x X_{\br}T_x^{-1}=X_{\br+\exx}$~(The action on the other Pauli operators are similarly defined, namely, a local Pauli operator is shifted by one lattice constant in the positive $x$-direction.). 
However, we come across an issue when we think of the action of the translation operator on the Gauss's laws. Indeed, we have
\begin{eqnarray}
    T_x G^{X}_{\br+\ex} T_x^{-1}\neq G^{X}_{\br+\frac{3\bf{e}_x}{2}} ,\quad  T_x G^{Z}_{\br+\ex} T_x^{-1}\neq G^{Z}_{\br+\frac{3\bf{e}_x}{2}} ,\quad  T_x G^{Z}_{\br+\ey} T_x^{-1}\neq G^{Z}_{\br+\frac{\bf{e}_x}{2}+\exx}.
\end{eqnarray}
Similar to the previous argument in the 1D case [see below~\eqref{dipole0}], the translational operator in the $x$-direction is not gauge invariant, whereas by the similar argument, one can verify that the translation operator in the $y$-direction is gauge invariant. 
To construct well-defined translational operator, we modify the operator as
\begin{eqnarray}
    T_x\to T_x W\vcentcolon=\mathcal{T}_x, \quad W\vcentcolon=(CZ) D (CZ).\label{51}
\end{eqnarray}
Here, 
\begin{eqnarray}
    CZ\vcentcolon=\prod_{\br}e^{\frac{\pi i}{4}(1-\sigma^Z_{\br})\left\{(1-\tau^Z_{\br})+(1-\tau^Z_{\br+\exx})+\left(1-\mu_{\br+\ex+\ey}\right)+\left(1-\mu_{\br+\ex-\ey}\right)\right\}},\label{35}
\end{eqnarray}
which corresponds to product of the controlled gate phase operators. 
Also, $D$ describes generalized duality map operator.
To give concise form of $D$, we introduce states
\begin{equation*}
    \ket{\{\tau_\br\},\{\sigma_\br\},\{\mu_\bp\}}\vcentcolon=\otimes_{\br}\ket{\tau_\br}\otimes_{\br}\ket{\sigma_\br}\otimes_{\bp}\ket{\mu_{\bp}}
    \quad(\tau_\br,\sigma_\br,\mu_\bp=0,1 \mod 2)
\end{equation*}
which are product of diagonal basis of the Pauli $Z$-operators on entire nodes and plaquettes. 
To wit,
\begin{eqnarray}
   \tau^Z_{\br_0}\ket{\{\tau_\br\},\{\sigma_\br\},\{\mu_\bp\}}&=&(-1)^{\tau_{\br_0}}\ket{\{\tau_\br\},\{\sigma_\br\},\{\mu_\bp\}},\nonumber\\
    \tau^X_{\br_0}\ket{\{\tau_\br\},\{\sigma_\br\},\{\mu_\bp\}}&=&\ket{\{\tau_{\br\neq\br_0};\tau_{\br_0}+1\},\{\sigma_\br\},\{\mu_\bp\}}.
\end{eqnarray}
The action of other Pauli operators, 
$\sigma^{X/Z}_{\br_0}$, $\mu^{X/Z}_{\bp_0}$
on the state is similarly defined. Using this state, the operator~$D$ is introduced as
\begin{eqnarray}
    D\vcentcolon= \frac{1}{\sqrt{2}}\sum_{\substack{\{\tau_{\br}\},\{\sigma_{\br}\},\{\mu_{\bp}\} \\ \{\tau^\prime_{\br}\},\{\sigma^\prime_{\br}\},\{\mu^\prime_{\bp}\} }}\Omega\ket{\{\tau^\prime_\br\},\{\sigma^\prime_\br\},\{\mu^\prime_\bp\}}\bra{\{\tau_\br\},\{\sigma_\br\},\{\mu_\bp\}},\label{D}
\end{eqnarray}
where a phase factor $\Omega$ is given by
\begin{eqnarray}
    \Omega=(-1)^{\sum_{\br}\left(\tau^\prime_{\br+\exx}+\tau^\prime_{\br}+\mu^\prime_{\br+\ex+\ey}+\mu^\prime_{\br+\ex-\ey}\right)\sigma_{\br}+(\sigma^\prime_{\br-\exx}+\sigma^\prime_{\br})\tau_\br+(\sigma^\prime_{\br+\eyy}+\sigma^\prime_{\br})\mu_{\br+\ex+\ey}}.\label{phasefactor}
\end{eqnarray}
The operator~$D$ is the generalization of the Kramers-Wannier duality operator~(see e.g.,~Refs.~\cite{yan2024generalized,Cao:2024qjj} for a relevant discussion in the case of one dimension) which is referred to as the \textit{bilinear phase map}. After some algebra, one has
\begin{eqnarray}
D\sigma^X_{\br}&=&\tau^Z_{\br+\exx}\tau^Z_{\br}\mu^Z_{\br+\ex+\ey}\mu^Z_{\br+\ex-\ey}D,\quad D\tau^Z_{\br+\exx}\tau^Z_{\br}\mu^Z_{\br+\ex+\ey}\mu^Z_{\br+\ex-\ey}=\sigma^X_{\br}D\nonumber\\
D\tau^X_{\br}&=&\sigma^Z_{\br-\exx}\sigma^Z_\br D,\quad D\sigma^Z_{\br-\exx}\sigma^Z_\br=\tau^X_{\br}D,\nonumber\\
D\mu^X_{\bp}&=&\sigma^Z_{\br+\eyy}\sigma^Z_{\br}D,\quad D\sigma^Z_{\br+\eyy}\sigma^Z_{\br}=\mu^Z_{\bp}D.\label{56}
\end{eqnarray}
See Appendix.~\ref{app22} for derivation.
Relations in~\eqref{56} indicate that $D$ gives 
isomorphisms between algebras of symmetric local operators induced by duality~\cite{Cobanera:2011wn}, which is implemented by gauging~$\mathbb{Z}_2$ $0$-form and $1$-form global symmetries in $(2+1)d$. 
%
\par
After defining operators $CZ$ and $D$, we now show the modified translation operator is gauge invariant. 
The operator~$W$ which is attached with the lattice translation operator~\eqref{51} is interpreted as the so-called “TST transformation’’  that maps between a spontaneous symmetry broken (SSB) state and the generalized cluster state protected by $\mathbb{Z}_2^{(0)}\times \mathbb{Z}_{2}^{(1)}$ symmetry~\footnote{In what follows, the upper index ``$(p)$'' in $\mathbb{Z}_N^{(p)}$ denotes form of the symmetry.}. Indeed, after some algebra, one finds (see also Fig.~\ref{tst})
\begin{eqnarray}
W\tau^Z_{\br+\exx}\tau^Z_{\br}\mu^Z_{\br+\ex+\ey}\mu^Z_{\br+\ex-\ey}&=&\tau^Z_{\br+\exx}\tau^Z_{\br}\mu^Z_{\br+\ex+\ey}\mu^Z_{\br+\ex-\ey}\sigma^X_{\br}W,\nonumber\\
\quad W\tau^Z_{\br+\exx}\tau^Z_{\br}\mu^Z_{\br+\ex+\ey}\mu^Z_{\br+\ex-\ey}\sigma^X_{\br}&=&\tau^Z_{\br+\exx}\tau^Z_{\br}\mu^Z_{\br+\ex+\ey}\mu^Z_{\br+\ex-\ey}W
,\nonumber\\
W\sigma^Z_{\br-\exx}\sigma^Z_{\br}=\sigma^Z_{\br-\exx}\sigma^Z_{\br}\tau^X_\br W
,&\quad& 
W\sigma^Z_{\br-\exx}\sigma^Z_{\br}\tau^X_\br=\sigma^Z_{\br-\exx}\sigma^Z_{\br}W
,\nonumber\\
W\sigma^Z_{\br+\eyy}\sigma^Z_{\br}
=
\sigma^Z_{\br+\eyy}\sigma^Z_{\br}\mu^X_\bp W
,&\quad& 
W\sigma^Z_{\br+\eyy}\sigma^Z_{\br}\mu^X_\bp
=
\sigma^Z_{\br+\eyy}\sigma^Z_{\br}W
,\nonumber\\
W\sigma^X_{\br}=\sigma_\br^XW,\quad W\tau^X_{\br}&=&\tau_\br^XW, \quad W\mu^X_{\bp}=\mu_\bp^XW.
\end{eqnarray}
It is worth mentioning that the fact that the translation operator is dressed with the TST transformation is not accidental; rather, it is an inevitable consequence of the underlying anomaly and the TQFT description, which we will turn to in the next section~(Sec.~\ref{s42}). \par
One can verify that the modified translation operator~\eqref{51} is gauge invariant. 
It follows that 
\begin{eqnarray}
    \mathcal{T}_xG^X_{\br+\ex}
    &=&
   { T}_x \left[X_{\br+\exx}\left(\sigma^X_{\br}\right)^{\hx+1} W\left(\sigma^X_{\br}\tau^Z_{\br+\exx}\tau^Z_{\br}\mu^Z_{\br+\ex+\ey}\mu^Z_{\br+\ex-\ey}\right)\right]
    \nonumber\\
    &=&
    {T}_x \left[ X_{\br+\exx}\left(\sigma^X_{\br}\right)^{\hx+1}\tau^Z_{\br+\exx}\tau^Z_{\br}\mu^Z_{\br+\ex+\ey}\mu^Z_{\br+\ex-\ey}\right]W
    =
    G^X_{\br+\frac{3\bf{e}_x}{2}}\mathcal{T}_x.
\end{eqnarray}
Similar line of thought leads to $\mathcal{T}_xG^{Z}_{\br+\ex} = G^{Z}_{\br+\frac{3\bf{e}_x}{2}} \mathcal{T}_x$ and
$ \mathcal{T}_x G^{Z}_{\br+\ey} =G^{Z}_{\br+\frac{\bf{e}_x}{2}+\exx} \mathcal{T}_x$, 
implying that lattice translation is now gauge invariant operator. 
Introducing passive lattice translation (i.e., lattice translation in the negative direction) as $\mathcal{T}_x^{-1}\vcentcolon=WT_x^{-1}$, moreover, 
we have the following fusion rule (see Appendix.~\ref{app22} for derivation):
\begin{eqnarray}
   \mathcal{T}_x\times \mathcal{T}_x^{-1}=(1+\eta^{(0)})\times\frac{1}{2} \prod_{a=x,y}(1+\eta^{(1)}_a)~\mathbb{I},\label{fusion2}
\end{eqnarray}
indicating that while the translation operator in $y$-direction is invertible, that is, $T_y\times T_y^{-1}=\mathbb{I}$, 
the one in the $x$-direction becomes non-invertible; the fusion rule between active and passive lattice translation 
is \emph{not} identity, rather, it
involves condensation defects, which are defects of dual symmetries~\eqref{loop}. \par
In summary, we introduce a 2D lattice model with the LSM anomaly~\eqref{lsm}, and gauge full internal symmetry. To construct well-defined translation operator, we modify the operator so that it is multiplied with an operator $W$, corresponding to the TST transformation. As a consequence, the modified translation operator becomes non-invertible. 
In Appendix.~\ref{3da}, we discuss three dimensional lattice model with the LSM anomaly and see that the similar non-invertible translation operator is obtained. Further, 
In the next section, we give complementary field theoretical argument to understand the non-invertible translation, highlighting how such an exotic translation operator comes up in view of anomaly inflow.

\section{Field theoretical interpretation}\label{sec4}
\begin{figure}
    \centering 
    \centering
\hspace*{-2.2cm}
\begin{tikzpicture}[scale=0.55, transform shape,
  font=\large,
  >=Latex,
  node distance=18mm and 30mm,
  box/.style={draw, thick, align=center, inner sep=7pt},
  arr/.style={-Latex, very thick, draw=blue!60},
  lab/.style={midway, fill=white, inner sep=2pt, font=\Large},
  purple/.style={box, fill=purple!15},
  bluebox/.style={box, fill=blue!12},
]

\node (top) {{\LARGE
$\frac{iN}{2\pi}\int_{M_{d+2}} A^{(p+1)}\wedge B^{(d-p)}\wedge e^{x}$
}};

\node[purple, below left=24mm and 62mm of top] (L)
{{\LARGE $(d\!-\!p\!-\!1)$-form dipole symmetry}};

\node[bluebox, below=28mm of top, text width=7.2cm] (C)
{{\Large Gauge full internal symmetries}};

\node[purple, below right=24mm and 62mm of top] (R)
{{\LARGE $p$-form dipole symmetry}};

\node[purple, below=43mm of C] (B)
{{\LARGE Non-invertible translation}};

\draw[arr] (top) -- node[lab, above, sloped] {Gauge $p$-form} (L);
\draw[very thick, draw=blue!60] (top) -- (C);
\draw[arr] (top) -- node[lab, above, sloped] {Gauge $(d\!-\!p\!-\!1)$-form} (R);

\draw[arr] (L) -- node[lab, below, sloped] {Gauge $(d\!-\!p\!-\!1)$-form dipole symmetry} (B);
\draw[arr] (C) -- (B);
\draw[arr] (R) -- node[lab, below, sloped] {Gauge $p$-form dipole symmetry} (B);

\end{tikzpicture} 
    \caption{Schematic overview of how different gauging procedures of the same
LSM anomaly~\eqref{eq:3dbulkanomaly} lead to qualitatively distinct symmetry structures.
Partial gauging of internal symmetries produces spatially modulated (dipole) symmetries,
whereas fully gauging all internal symmetries inevitably promotes lattice translation
to a non-invertible symmetry defect.
This correspondence is naturally understood from anomaly inflow, where the translation
defect must carry additional topological degrees of freedom to cancel the mixed anomaly. Here, ``gauge $(d-p-1)$- or $p$-form dipole symmetry'' means gauging spatially modulated part of the symmetry. 
}
    \label{diagram}
\end{figure}
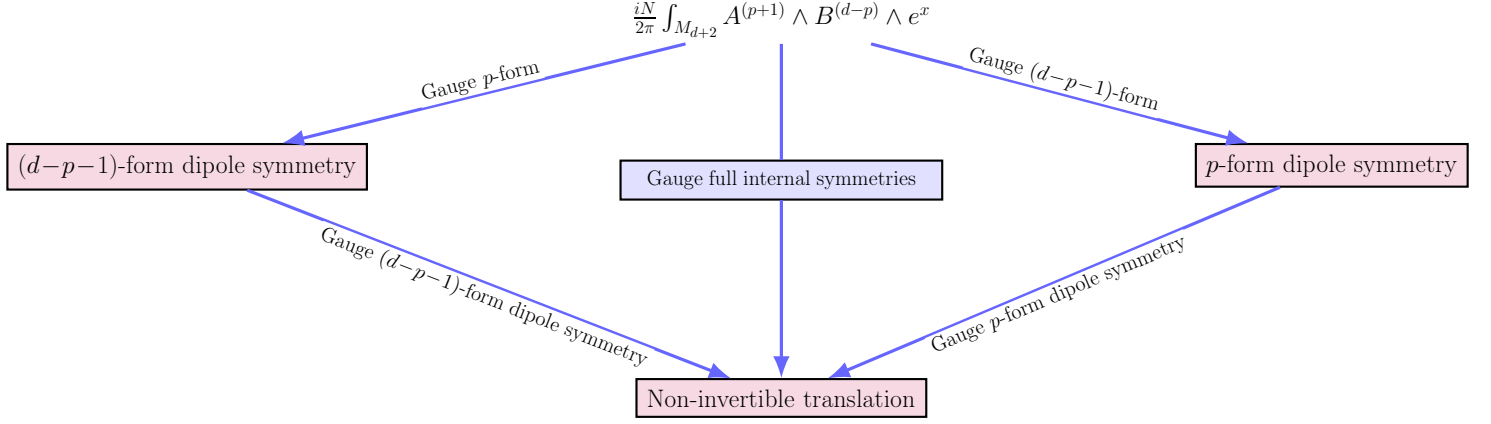

In this section, we provide field theoretical interpretations of our results~\footnote{While non-invertible translations were discussed in continuum field theories in Ref.~\cite{Seiberg:2024wgj,Seiberg:2024yig}, our field theoretical analysis emphasizes their origin from LSM anomalies.}. More precisely, we discuss how non-invertible translation symmetries emerge from the LSM anomaly in terms of anomaly inflow framework. Here, we consider LSM anomalies in general spatial dimensions, including those discussed in Sec.~\ref{sec2}, Sec.~\ref{sec3}, and Appendix.~\ref{3da}. Generally, the system with the LSM anomaly involving two internal and one translational symmetries
is captured by counter-term defined in one dimension higher. Theories with such an anomaly counter-term contains rich physics. Indeed, it is discussed that 
modulated symmetries have been emerged  by gauging one of the internal symmetries. Non-invertible translation is obtained by gauging full internal symmetries. See also Fig.~\ref{diagram}.
For the sake of completeness, in what follows, we first give overview of the emergence of the modulated symmetries via gauging one of the internal symmetries and then discuss how non-invertible translational symmetries comes into play when gauging full internal symmetries, which complies well with our analysis on the lattice model. 
\subsection{Recap:~LSM anomaly and modulated symmetry}
The LSM anomaly involving internal $\mathbb Z_N$
$p$-and $(d-p)$-form 
 symmetries and lattice translation symmetry in the $x$-direction in $(d+1)$-spacetime dimension
is captured by the following anomaly counter-term described defined in
one dimension higher, that is, the LSM anomaly is described by
\begin{equation}\label{eq:3dbulkanomaly}
    \frac{iN}{2\pi}\int_{M_{d+2}}A^{(p+1)}\wedge B^{(d-p)}\wedge e^x,
\end{equation}
Here, the $1$-form field $e^x$ is defined as $e^x\vcentcolon=dx$ which is referred to as the \emph{foliation field} along which $(d+1)$-dimensional submanifold is stacked
and $M_{d+2}$ denotes $(d+2)$-dimensional spacetime manifold. Also, $A^{(p+1)},~B^{(d-p)}$ denote $(p+1)$-form and $(d-p)$-form background $\mathbb Z_N$ gauge field, respectively. 
Physically, the theory~\eqref{eq:3dbulkanomaly} is interpreted as 
 weak SPTs protected by 
 $\mathbb{Z}_N^{(p)}\times\mathbb{Z}_{N}^{(d-p-1)}$
 symmetry, stacked along $x$-direction~\cite{Ebisu:2023idd,Pace:2025hpb,2024multipole,Antinucci:2025fjp}. 
 
To see the emergence of the dipole symmetry, we gauge the $\mathbb Z_N$ symmetry with gauge field~$A^{(p+1)}$ by promoting it to dynamical gauge field $a^{(p+1)}$ and coupling to the dual background gauge field $\tilde{A}^{(d-p)}$. The gauge symmetry should be free of anomaly, meaning we need to couple
\begin{equation}
    \frac{iN}{2\pi}\int_{\partial M_{d+2}} a^{(p+1)}\wedge \tilde{A}^{(d-p)}=\frac{iN}{2\pi}\int_{M_{d+2}} d(a^{(p+1)}\wedge \tilde{A}^{(d-p)})=\frac{iN}{2\pi}\int_{M_{d+2}} (-1)^{p+1}a^{(p+1)}\wedge d\tilde{A}^{(d-p)},
\end{equation}
at the boundary to cancel the bulk anomaly term \eqref{eq:3dbulkanomaly}. Here, we have used the flatness condition $da^{(p+1)}=0$ for the~$\mathbb Z_N$ gauge field $a^{(p+1)}$ in the last equality. Hence, we have
\begin{equation}
    a^{(p+1)}\wedge B^{(d-p)}\wedge e^x +(-1)^{p+1}a^{(p+1)}\wedge d \tilde{A}^{(d-p)}=0.
\end{equation}
This yields the modified flatness condition for the dual gauge field, namely, 
\begin{equation}
    d\tilde{A}^{(d-p)}=(-1)^{p}B^{(d-p)}\wedge e^x.
\end{equation}
Together with $dB^{(d-p)}=0$, the gauged theory exhibits the same flatness condition the case of the $(d-p-1)$-form dipole symmetry~\cite{Ebisu:2023idd}. 
\par
If we start with the same LSM anomaly and perform gauging the other subgroup, i.e., gauging $(d-p-1)$-form symmetry, by the similar argument, we obtain the $p$-form dipole symmetry.


 \subsection{Non-invertible translation symmetry}\label{s42}
%
When all internal symmetries are fully gauged, the anomaly can no longer
be absorbed into a modulated action of internal symmetries.
Instead, lattice translation itself must be modified.
As a consequence, the bare translation defect necessarily carries an anomalous phase.
To cancel this anomaly, additional topological degrees of freedom must live on the
translation defect, rendering it intrinsically non-invertible.
From this perspective, non-invertible translation is not an accidental lattice artifact,
rather, an inevitable outcome of anomaly inflow, and is naturally identified with an
$N$-ality defect whose grading and fusion rules are fixed by the original LSM anomaly.
The central logic is illustrated in Fig.~\ref{diagram}.\par
%
%
%
To see this point more clearly, 
we simplify the problem by
regarding the translation symmetry as an internal $0$-form $\mathbb{Z}_{L_x}$ ($L_x$:~system size in the $x$-direction) symmetry, in accordance with the crystalline equivalence principle~\cite{Thorngren:2016hdm, Else:2018eas}, where crystalline symmetries can be regarded as internal symmetries. We assume $L_x$ to be a multiple of $N$, and write $L_x=N\ell$ . In this setup, the anomaly inflow action \eqref{eq:3dbulkanomaly} can be interpreted as
\begin{equation}
    \frac{2\pi i}{N}\int_{M_{d+2}}A^{(p+1)} \cup B^{(d-p)} \cup C^{(1)}, \label{type3}
\end{equation}
where $C^{(1)}$ is a background $\mathbb{Z}_{L_x}$ gauge field. Here, we have rewritten the expression from differential-form language into the one in terms of cochain. We denote the original theory with the LSM anomaly and the theory obtained after gauging full internal symmetries (other than translation symmetry) by $\mathcal{\chi}$ and $\widehat{\mathcal{\chi}}$, respectively, which are defined on 
 $(d+1)$-dimensional spacetime manifold $X_{d+1}$.
The partition function of the gauged theory $\widehat{\mathcal{\chi}}$ is given by~\footnote{The normalization factor is omitted here.}
\begin{equation}
    Z_{\widehat{\mathcal{\chi}}}[\hat{A}^{(d-p)},\hat{B}^{(p+1)}] \sim \sum_{\substack{a^{(p+1)}\in H^{p+1}(X_{d+1},\mathbb{Z}_N) \\ b^{(d-p)}\in H^{d-p}(X_{d+1},\mathbb{Z}_N)}} Z_{\mathcal{\chi}}[a^{(p+1)}, b^{(d-p)}]\ e^{\frac{2\pi i}{N}\int a^{(p+1)}\cup \hat{A}^{(d-p)} + b^{(d-p)}\cup \hat{B}^{(p+1)}},
\end{equation}
where $\hat{A}^{(d-p)}$ and $\hat{B}^{(p+1)}$ represent background gauge fields for dual $(d-p-1)$-form and $p$-form $\mathbb{Z}_N$ symmetries, respectively. In what follows, all summations are taken over cohomology classes, and the summation indices will be omitted.
For convenience, we define two topological manipulations~S and T on a theory with $\mathbb{Z}_N^{(p)}\times\mathbb{Z}_N^{(d-p-1)}$ symmetry as
\begin{equation}
    \begin{aligned}
        \mathrm{S} : Z[A^{(p+1)}, B^{(d-p)}] &\longrightarrow \#\sum_{a^{(p+1)},b^{(d-p)}} Z[a^{(p+1)}, b^{(d-p)}]\ e^{\frac{2\pi i}{N}\int a^{(p+1)}\cup \hat{A}^{(d-p)} + b^{(d-p)}\cup \hat{B}^{(p+1)}} \\
        \mathrm{T} : Z[A^{(p+1)}, B^{(d-p)}] &\longrightarrow Z[A^{(p+1)}, B^{(d-p)}]\ e^{-\frac{2\pi i}{N}\int A^{(p+1)} \cup B^{(d-p)}},
    \end{aligned}\label{manipu}
\end{equation}
where $\#$ denotes a normalization factor depending on the topology of spacetime manifold. The manipulations 
S and T, which implement gauging of $\mathbb{Z}_N^{(p)}\times\mathbb{Z}_N^{(d-p-1)}$ symmetry and stacking of $\mathbb{Z}_N^{(p)}\times\mathbb{Z}_N^{(d-p-1)}$ SPT, respectively, are subject to the relations $\mathrm{S}^2=1$ and $\mathrm{T}^N=1$. 

Translation symmetry is constructed from a combination of the topological manipulations~\eqref{manipu},
$\mathrm{STS}$\footnote{For $N=2$, the relation $\mathrm{STS}=\mathrm{TST}$.}. Indeed, one can check that the partition function of the gauged theory $\widehat{\mathcal{\chi}}$ is invariant under the composite operation of STS and symmetry transformation for $\mathbb{Z}_{L_x}$. Namely, 
\begin{equation}
    Z_{g\mathrm{STS}\widehat{\mathcal{\chi}}}[\hat{A}^{(d-p)},\hat{B}^{(p+1)}] = Z_{\widehat{\mathcal{\chi}}}[\hat{A}^{(d-p)},\hat{B}^{(p+1)}],\label{1} 
\end{equation}
where $g$ is a generator of $\mathbb{Z}_{L_x}$ symmetry and acts on the partition function of the original theory~$\mathcal{\chi}$ as
\begin{equation} \label{sym action}
Z_{g\mathcal{\chi}}[A^{(p+1)}, B^{(d-p)}] = Z_{\chi}[A^{(p+1)}, B^{(d-p)}]\ e^{\frac{2\pi i}{N}\int A^{(p+1)} \cup B^{(d-p)}}
\end{equation}
due to the mixed anomaly \eqref{type3}. The two equations~\eqref{1}\eqref{sym action} lead to 
that $g$ alone is not a symmetry defect of the gauged theory and should be 
dressed with
STS, whereas $g^{\alpha N}(\alpha\in \mathbb{Z}_{\ell})$ remains a symmetry defect on its own. 
Therefore, the non-invertible translation defect is constructed from the composition of lattice translation defect $g$ and topological manipulation STS, and it has the~$\mathbb{Z}_N$-grading since $(\mathrm{STS})^N=1$. In fact, this corresponds exactly to \eqref{TST}, \eqref{51},  \eqref{STS}, and~\eqref{114} in the lattice description. 

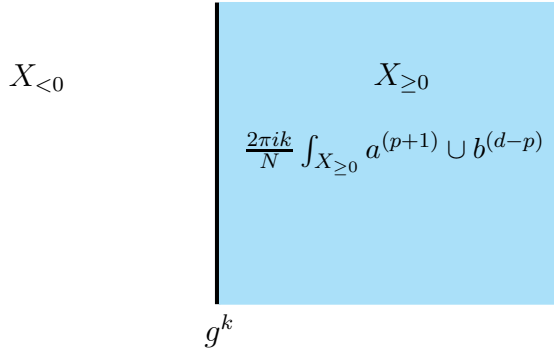
\begin{figure}
    \centering 
    \begin{tikzpicture}
    \coordinate (A) at (1,0);
    \coordinate (XL) at (-1.4,3);
    \coordinate (XR) at (3.4,3);

   \draw[white] (-3,0)--(-3,4);
   \draw[line width = 2][black] (0.975,0)--(0.975,4);
    
    \filldraw[cyan!30,opacity=0.7] (1,0) rectangle (5.5,4);
    
    \node at (3.3,2) {$\frac{2\pi ik}{N}\int_{X_{\ge 0}}a^{(p+1)} \cup b^{(d-p)}$};
    \node at (A) [below] {\large $g^k$};
    \node at (XL) {\large $X_{<0}$};
    \node at (XR) {\large $X_{\ge0}$};
\end{tikzpicture} 
    \caption{Anomaly inflow from the right bulk SPT phase to symmetry defect.}
    \label{defect}
\end{figure}

Once we understand that the translation operator, which is interpreted as the translational symmetry defect in TQFT perspective, is dressed with the topological manipulations, STS, we move on to discuss the fusion rule of the translation operator. To do so, we 
investigate the non-invertible defects~\footnote{Similar constructions of non-invertible defects have been studied in several works \cite{Tachikawa:2017gyf,Choi:2022jqy,Kaidi:2021xfk, Kaidi:2023maf}.}. To construct the non-invertible defects, we consider placing a symmetry defect $g^k$ on a codimension-$1$ surface $Y$. 
Due to the mixed anomaly \eqref{type3}, the SPT phase 
\begin{equation}
\frac{2\pi ik}{N}\int_{X_{\ge 0}}a^{(p+1)} \cup b^{(d-p)}
\end{equation}
arises 
on the half-space $X_{\ge 0}$ adjacent to a symmetry defect $g^k$ (see Fig. ~\ref{defect}).
To promote $g^k$ to a symmetry defect of the gauged theory, we consider coupling a TQFT on the defect $g^k$, which cancels the anomaly from the SPT phase\footnote{The choice of such a TQFT is not unique. However, it is known that such TQFTs can be decomposed into the product of the \emph{minimal} TQFT associated with the anomaly and a decoupled TQFT \cite{Hsin:2018vcg}.}. 
This is in line with the fact that 
to cancel the anomaly from the SPT phase, one generally needs d.o.f in higher dimensions rather than one-dimensional ones. 
%
As a consequence, 
the presence of this TQFT makes the defect non-invertible. 

Based on this argument, we explicitly 
demonstrate
the non-invertible defect and compute their fusion rules in the case of $N=2$. 
The SPT phase attached to the symmetry defect $g$ is given by
\begin{equation}
\frac{2\pi i}{2}\int_{X_{\ge 0}}a^{(p+1)} \cup b^{(d-p)}.
\end{equation}
To cancel the anomaly from this SPT phase, we couple the symmetry defect with $d$-dimensional TQFT\footnote{We have omitted the normalization factor.}
\begin{eqnarray}
     \sum_{\substack{\phi_a^{(p)}\in C^{p}(Y,\mathbb{Z}_2) \\ \phi_b^{(d-p-1)}\in C^{d-p-1}(Y,\mathbb{Z}_2)}} \exp\Bigg[\frac{2\pi i}{2} \int_Y\phi_b^{(d-p-1)}a^{(p+1)}+\phi_a^{(p)}b^{(d-p)}+\phi_a^{(p)}\delta\phi_b^{(d-p-1)}\Bigg],
\end{eqnarray}
where $\phi_a^{(p)}$ and $\phi_b^{(d-p-1)}$ are $\mathbb{Z}_2$-valued $p$- and $(d-p-1)$- cochains supported on $Y$, respectively. Indeed, one can check that under the gauge transformation
\begin{equation}
    \begin{cases}
        a^{(p+1)} \rightarrow a^{(p+1)} + \delta\alpha^{(p)} \\
        b^{(d-p)} \rightarrow b^{(d-p)} + \delta\beta^{(d-p-1)} \\
        \phi_a^{(p)} \rightarrow \phi_a^{(p)} + \alpha^{(p)} + \delta\tilde{\alpha}^{(p-1)} \\
        \phi_b^{(d-p-1)} \rightarrow \phi_b^{(d-p-1)} + \beta^{(d-p-1)} + \delta\tilde{\beta}^{(d-p-2)},
    \end{cases}
\end{equation}
the gauge variation of the TQFT cancels the anomaly from the bulk SPT phase. Therefore, up to an overall normalization factor, the non-invertible defect is defined as
\begin{eqnarray}
    \mathcal{N}(Y) \sim \sum_{\substack{\phi_a^{(p)}\in C^{p}(Y,\mathbb{Z}_2) \\ \phi_b^{(d-p-1)}\in C^{d-p-1}(Y,\mathbb{Z}_2)}} g(Y) \exp\Bigg[\frac{2\pi i}{2} \int_Y\phi_b^{(d-p-1)}a^{(p+1)}+\phi_a^{(p)}b^{(d-p)}+\phi_a^{(p)}\delta\phi_b^{(d-p-1)}\Bigg].
\end{eqnarray}
The fusion rules are given by\footnote{Upon specifying the normalization factor, all $\sim$ can be replaced by $=$. }
\begin{equation} \label{fusion in field theory}
    \begin{aligned}
        \mathcal{N} \times \eta_a &= \eta_a \times \mathcal{N} = \mathcal{N} \\
        \mathcal{N} \times \eta_b &= \eta_b \times \mathcal{N} = \mathcal{N} \\
        \mathcal{N} \times \mathcal{N}^{\dagger} & \sim C_aC_b,
    \end{aligned}
\end{equation}
where $\eta_a$ and $\eta_b$ are Wilson surface of $a^{(p+1)}$ and $b^{(d-p)}$. To wit,
\begin{equation}
    \eta_a =\exp\Big(i\pi\oint a^{(p+1)}\Big), \quad \eta_b =\exp\Big(i\pi\oint b^{(d-p)}\Big).
\end{equation}
These operators generate dual symmetry $\mathbb{Z}_2^{(d-p-1)} \times \mathbb{Z}_2^{(p)}$. Moreover, $C_a$ and $C_b$ represent the condensation defects, associated with $\eta_a$ and $\eta_b$, respectively, which are written as
\begin{equation}
    C_a =  \frac{1}{|H^0(Y, \mathbb{Z}_2)|}\sum_{\Sigma_{p+1}\in H_{p+1}(Y, \mathbb{Z}_2)} \eta_a(\Sigma_{p+1}), \quad  C_b = \frac{1}{|H^0(Y, \mathbb{Z}_2)|}\sum_{\Sigma_{d-p}\in H_{d-p}(Y, \mathbb{Z}_2)} \eta_b(\Sigma_{d-p}).
\end{equation}
These are summation of symmetry defects over the distinct homology classes. 
A detailed derivation of the fusion rules~\eqref{fusion in field theory} is provided in Appendix.~\ref{appendix.E}. These fusion rules are consistent with those obtained in the lattice description. In fact, choosing the spatial manifold as $Y = T^d$ yields precisely the same fusion rules [see~\eqref{fusion},\eqref{fusion2},\eqref{conde}].

We close this subsection by making a few remarks on this symmetry category for general~$N$. The symmetry category discussed here is mathematically described by $\mathbb{Z}_N$-graded fusion category\footnote{The $A$-graded fusion category $\mathcal{C}$ is a fusion category which admits a direct sum decomposition $\mathcal{C}=\bigoplus_{g\in A}\mathcal{C}_g$, where the tensor product is defined as $\otimes : \mathcal{C}_g\times \mathcal{C}_h\rightarrow\mathcal{C}_{gh}$, $\forall g,h\in A$, where $A$ is a finite group.} \cite{Etingof:2009yvg, Gelaki:2009blp}, since the symmetry defects carry $\mathbb{Z}_N$-grading, and it is referred to as the $N$-ality symmetry in the context of physics. In particular, for $d=1$, the symmetry category coincides with $N$-ality symmetry studied in Refs.\cite{Maeda:2025rxc}, up to the invertible defect $g^N$ in this section, which is equivalently $N$-site lattice translation operator in lattice description. The fusion rule is given by
\begin{equation}
    \mathcal{N}_k \times \mathcal{N}_{k'} =\frac{dd'D}{\ell^2}\left(\sum_{i,j=0}^{\ell/D-1}\eta_a^{\frac{N}{\ell}i}\eta_b^{\frac{N}{\ell}j}\right)\mathcal{N}_{k+k'}, \label{fusion rule}
\end{equation}
where $\ell=\mathrm{lcm}(d,d')$ and $D=\frac{N}{\mathrm{gcd}(N,k+k')}$. We note that when $k+k'=N$, we regard $\mathcal{N}_{k+k'}$ as $g^N$, which corresponds to $N$-site lattice translation operator $T_x^N$ in lattice description. See Refs.\cite{Maeda:2025rxc} for the definition of $\mathcal{N}_k$ and the derivation of fusion rules. Indeed, as shown in the Appendix.~\ref{appendix.A}, the non-invertible translation operators satisfy the above fusion rules for $d=1$ and $N=p$, where~$p$ is prime.

\section{Outlook}\label{sec5}
In this work, we have shown that gauging the full internal symmetry in the presence of an LSM anomaly leads to non-invertible lattice translation operators. Together with previous results on spatially modulated symmetries obtained from partial gauging, our analysis completes a coherent picture in which different gauging procedures of the same internal symmetry give rise to distinct but related manifestations of the underlying anomaly.

One can also study LSM system with the type-IV anomaly, involving two internal and two translational symmetries. It has been shown that 
gauging a subgroup of the internal symmetries leads to 
a modulated symmetry consisting of both $1$-form and $0$-form components~\cite{oishi2026type,Ebisu:2025mtb}. The structure of this modulated symmetry depends crucially on the system size, in particular its parity, 
reflecting the role of symmetry defects under the crystalline equivalence principle. 
From this perspective, it is
natural to expect that 
fully gauging internal symmetries generically promotes lattice translations to non-invertible also in the type-IV case. This suggest that the emergence of non-invertible translation in a universal consequence of mixed crystalline anomalies, beyond the type-III setting studied in this work. 
However, compared to the type-III case, the structure of the Gauss law becomes more involved, making an explicit construction technically more subtle. We leave a detailed analysis of this type-IV case for future work.

An immediate direction for future work is to generalize our construction to more general crystalline symmetries beyond lattice translations. While we have focused on translational symmetry, LSM anomalies involving rotations, reflections, or non-symmorphic symmetries are known to impose similarly strong constraints on lattice systems~\cite{Thorngren:2016hdm, Else:2018eas}. It would be highly interesting to determine whether gauging the full internal symmetry in such settings likewise produces non-invertible crystalline symmetry operators, and how their defect fusion rules encode the corresponding anomaly data.

Another important avenue concerns the formulation of our results within the framework of symmetry topological field theories (SymTFTs)~\cite{Pace:2025hpb,Antinucci:2025fjp,Yao:2025iia,Cao:2023rrb}. Non-invertible translation operators naturally define a fusion category of symmetry defects, suggesting that the LSM anomaly may admit a purely topological characterization in terms of a bulk SymTFT. Developing such a formulation could provide a systematic classification of non-invertible crystalline symmetries induced by anomalies, and clarify their relation to recent developments in non-invertible symmetries and categorical symmetry structures.

Studying emergence of more general multipole symmetries is an another natural direction. To realize a quadrupole-like structure, one may consider a model, which consists of four spin operators arranged in a plaquette configuration, jointly with an appropriate subsystem symmetry. Alternatively, one can introduce multiple spin degrees of freedom on each lattice site such that each planar subsystem exhibits an LSM-type constraint. In such constructions, gauging suitable subgroups of the internal symmetry may lead to emergent subsystem symmetries or multipole conservation laws, potentially realizing quadrupole-type symmetries.  A systematic construction of such models and their associated non-invertible translation operators remains an interesting direction for future work.

Finally, our results suggest a close connection between non-invertible translations and foliated or fracton phases of matter. Since such phases are characterized by subdimensional mobility and foliation-dependent structures, it is natural to ask whether non-invertible crystalline symmetries provide a unifying language for describing mobility constraints and anomaly induced obstructions in these systems. %
\section*{Acknowledgement}
We thank Weiguang Cao, Bo Han, Masazumi Honda, Ho Tat Lam, Shang-Qiang Ning, Jun Maeda, Taiichi Nakanishi, Salvatore D. Pace and Soichiro Shimamori for discussion. This work is in part supported by JST CREST (Grant No.~JPMJCR24I3), JST SPRING (Grant No.~JPMJSP2110).

\begin{appendix}
\section{Non-invertible translation symmetry for $\mathbb{Z}_p$} \label{appendix.A}
In this appendix, we generalize the discussion of Sec.~\ref{sec2} to prime $N=p$. We give an explicit expression for the non-invertible translation operator and use it to derive the fusion rules. The discussion here largely parallels that of Sec.~\ref{sec2}; however, the construction of the non-invertible translation operators and their fusion rules are slightly modified.

\subsection{Construction of non-invertible translation operators}
Let us gauge the dipole symmetry generated by $Q_x$. As in the Sec.~\ref{sec2}, we introduce extended Hilbert space on each vertex, and gauge variables $\m{X}_j$ and $\m{Z}_j$ acting on each vertex. 
We define the Gauss's law operator as
\begin{equation}
    G_{(j.j+1)}=\m{X}_j\big(\s^X_{(j,j+1)}\big)^j\m{X}_{j+1}^\dagger, 
\end{equation}
and impose the Gauss's law constraint, namely, $G_{(j.j+1)}=1$ at each link. 
By minimally coupling to \eqref{dipole Hamiltonian}, we obtain the gauge invariant Hamiltonian
\begin{equation}
    H_{\mathrm{XZ}/\mathbb{Z}_p\times\mathbb{Z}_p} = \sum_{j=1}^L \big(J_\mathrm{X}\s^X_{(j,j+1)} + J_\mathrm{Z}\s^{Z\dagger}_{(j-1,j)}\m{Z}_j^{-j+1}(\s^{Z}_{(j,j+1)})^2\m{Z}_{j+1}^{j+1}\s^{Z\dagger}_{(j+1,j+2)}\big) + h.c.. \label{gauged dipole Hamiltonian}
\end{equation}
One can readily find that this model has two $\mathbb{Z}_p$ $0$-form symmetries generated by
\begin{equation}
    \widehat{U}_X=\prod_{j=1}^{L}\m{Z}_j, \quad \widehat{U}_Z=\prod_{j=1}^{L}\s^X_{(j,j+1)}. \label{Z_N sym}
\end{equation}
However, this Hamiltonian does not possess the lattice translation symmetry, since lattice translation operator is not gauge invariant\footnote{Note that $T_x^{p\alpha}(\alpha=1,2,\cdots, \ell-1)$ remain as ordinary lattice translation symmetry since they are gauge invariant, where the system size $L=p\ell$.}, 
\begin{equation}
    T_xG_{(j,j+1)}T_x^{-1}\ne G_{(j+1,j+2)}.
\end{equation}
To make translation operator gauge invariant, we modify the operator as
\begin{equation}
    T_x\rightarrow \mathcal{T}_{x,1} := T_xW,\label{STS}
\end{equation}
where $W$ implements the transformation
\begin{equation}
    \begin{aligned} 
        \s^Z_{(j,j+1)}\s^{Z\dagger}_{(j+1,j+2)} &\rightarrow \s^Z_{(j,j+1)}\m{Z}_{j+1}\s^{Z\dagger}_{(j+1,j+2)} \\
        \m{X}_j\m{X}_{j+1}^\dagger &\rightarrow \m{X}_{j}\s^X_{(j,j+1)}\m{X}_{j+1}^\dagger
    \end{aligned}
\end{equation}
This transformation can be interpreted as the topological manipulation STS, where S and T represent the gauging of two $\mathbb{Z}_p$ 0-form symmetries \eqref{Z_N sym} and the stacking of SPT phase protected by the same symmetry, respectively. For $N=2$, the relation $\mathrm{STS}=\mathrm{TST}$ holds, and this corresponds to the so-called Kennedy-Tasaki transformation~\cite{Kennedy:1992ifl, Kennedy:1992tke, Oshikawa:1992smv, Li:2023ani}. Since $W$ involves a gauging operation, it is non-invertible. Therefore, the modified translation operator is also non-invertible.

To compute the fusion rule, we first give an alternative representation of the non-invertible translation operator. So far, we have rendered the translation operator gauge invariant by modifying it with $W$. Here, instead, we consider making it gauge invariant via minimal coupling. That is, we modify the translation operator \eqref{translation} as follows\footnote{More precisely, in order to obtain a physical operator, it is necessary to project onto the subspace with $G_{(j,j+1)}=1$.},
\begin{equation}
    \widetilde{T}_x=\prod_{j=1}^{L-1}\widetilde{T}_j, \quad \widetilde{T}_j=\frac{1}{N}\sum_{a,b=0}^{p-1}\omega^{ab}(\s^X_{(j,j+1)})^a\big(\s^{Z\dagger}_{(j-1,j)}\m{Z}_j^{-j+1}(\s^{Z}_{(j,j+1)})^2\m{Z}_{j+1}^{j+1}\s^{Z\dagger}_{(j+1,j+2)}\big)^b. \label{non-invertible translation}
\end{equation}
This operator acts on the gauge invariant local operators as\footnote{The Gauss's law implies $X_LX_1^\dagger=1$.}
\begin{equation}
    \begin{aligned}
        \widetilde{T}_x\s^X_{(j,j+1)} &= \s^X_{(j+1,j+2)} \widetilde{T}_x\quad (j\ne L-1,L), \\
        \widetilde{T}_xh_{j,j+1} &= h_{j+1,j+2} \widetilde{T}_x\quad 
        (j\ne L-1,L), \\
        \widetilde{T}_x\m{X}_j\m{X}_{j+1}^\dagger &= \m{X}_{j+1}\s^X_{(j+1,j+2)}\m{X}_{j+2}^\dagger\widetilde{T}_x \quad 
        (j\ne L),
    \end{aligned}
\end{equation}
where $h_{j,j+1}=\s^{Z\dagger}_{(j-1,j)}\m{Z}_j^{-j+1}(\s^{Z}_{(j,j+1)})^2\m{Z}_{j+1}^{j+1}\s^{Z\dagger}_{(j+1,j+2)}$. However, for the operator around site $L$, we find non-local operators involving $\widehat{U}_X$ and $\widehat{U}_Z$. Indeed, 
\begin{equation}
    \begin{aligned}
        \widetilde{T}_x\s^X_{(L,1)} &= \s^X_{(1,2)}\widehat{U}_Z\widetilde{T}_x,\quad
        \widetilde{T}_x\s^X_{(L-1,L)} = \s^X_{(L,1)}\widehat{U}_Z^{-1}\widetilde{T}_x,\\ \widetilde{T}_xh_{L,1} &= h_{1,2}\widehat{U}_X\widetilde{T}_x, \quad \widetilde{T}_xh_{L-1,L} = h_{L,1}\widehat{U}_X^{-1}\widetilde{T}_x.
    \end{aligned}
\end{equation}
To circumvent this issue, we multiply $\widetilde{T}_x$ with the condensation operator \cite{Roumpedakis:2022aik, Choi:2024rjm} for $\widehat{U}_X$ and $\widehat{U}_Z$, defined as 
\begin{equation}
    C_{X}\vcentcolon=\sum_{a=0}^{p-1}(\widehat{U}_X)^a, \quad C_{Z}\vcentcolon=\sum_{a=0}^{p-1}(\widehat{U}_Z)^a.
\end{equation}
Hence, the well-defined translation operator is given by
\begin{equation}
    \mathcal{T}_{x,1}:=\frac{C_{X}}{\sqrt{p}}\frac{C_{Z}}{\sqrt{p}}\widetilde{T}_x.
\end{equation}
However, due to the condensation operator, the translation symmetry becomes non-invertible.
The label indicates the grading of the non-invertible translation operator\footnote{The grading originates from the power of the lattice translation operator.}. As we will see later, in general, a non-invertible translation operator with grading $k\ (k=1,2,\cdots, p-1)$ can be obtained from $\widetilde{T}_x^k$. The non-invertible translation operator $\mathcal{T}_{x,1}$ acts on the gauge invariant local operators as
\begin{equation}
    \begin{aligned}
       \s^X_{(j,j+1)} \rightarrow \s^X_{(j+1,j+2)} ,\quad
        h_{j,j+1} \rightarrow h_{j+1,j+2}, \quad
        \m{X}_j\m{X}_{j+1}^\dagger \rightarrow \m{X}_{j+1}\s^X_{(j+1,j+2)}\m{X}_{j+2}^\dagger,
    \end{aligned}
\end{equation}
and thus commutes with Hamiltonian \eqref{gauged dipole Hamiltonian}. Its non-invertibility can also be seen from the fact that it has a non-trivial kernel,
\begin{equation}
    \widehat{U}_X\mathcal{T}_{x,1}=\mathcal{T}_{x,1}, \quad \widehat{U}_Z\mathcal{T}_{x,1}=\mathcal{T}_{x,1}.
\end{equation}

The non-invertible translation operator with grading $k$ can be constructed in the same way and is given by 
\begin{equation} \label{non-inv with k}
    \mathcal{T}_{x,k}:=\frac{C_{X}}{\sqrt{p}}\frac{C_{Z}}{\sqrt{p}}\widetilde{T}_x^k.
\end{equation}
This operator is gauge invariant and acts on the gauge invariant local operators as
\begin{equation}
   \begin{aligned}
       \s^X_{(j,j+1)} \rightarrow \s^X_{(j+k,j+k+1)} ,\quad
        h_{(j,j+1)} \rightarrow h_{(j+k,j+k+1)}, \quad
        \m{X}_j\m{X}_{j+1}^\dagger \rightarrow \m{X}_{j+k}(\s^X_{(j+k,j+k+1)})^k\m{X}_{j+k+1}^\dagger,
    \end{aligned}
\end{equation}
This action implies that $\mathcal{T}_{x,k}$ commutes with the Hamiltonian \eqref{gauged dipole Hamiltonian}. 

\subsection{Fusion rules} \label{appendix.A.2}
We compute the fusion rules of non-invertible translation symmetry using expression \eqref{non-inv with k}. 

The fusion rule of $\mathcal{T}_{x,k} \times \mathcal{T}_{x,k'}$ can be computed explicitly as follows:
\begin{equation}
    \begin{aligned}
        \mathcal{T}_{x,k} \times \mathcal{T}_{x,k'} &= 
        \frac{C_X}{\sqrt{p}}\frac{C_Z}{\sqrt{p}}\widetilde{T}_x^k\times \frac{C_X}{\sqrt{p}}\frac{C_Z}{\sqrt{p}}\widetilde{T}_x^{k'} \\
        &= \frac{C_X^2}{p}\frac{C_Z^2}{p}\widetilde{T}_x^{k+k'} \\
        &=p \frac{C_X}{\sqrt{p}}\frac{C_Z}{\sqrt{p}}\widetilde{T}_x^{k+k'}     
    \end{aligned}
\end{equation}
where we have used $C_X^2=pC_X$, $C_Z^2=pC_Z$ and the fact that $\widetilde{T}_x$ commutes with the condensation operators $C_X, C_Z$. When $k+k'=p$, $\widetilde{T}_x^{k+k'}$ has a same action on the gauge invariant local operators as the $N$-site lattice translation $T_x^p$ up to $\widehat{U}_X, \widehat{U}_Z$. However, $\widehat{U}_X$ and $\widehat{U}_Z$ can be absorbed into the condensation operators. Therefore, we obtain the following fusion rule,
\begin{equation}
    \mathcal{T}_{x,k} \times \mathcal{T}_{x,k'} =
    \begin{cases}
        C_XC_ZT_x^p & (k+k'=p) \\
        p\mathcal{T}_{x,k+k'} & (\mathrm{otherwise}).
    \end{cases}
\end{equation}
This result is consistent with the field-theoretical description discussed in Sec.~\ref{sec4}, where $\mathcal{T}_{x,k}$ is replaced by $\mathcal{N}_k$.

\section{Dipole symmetry in 2D}\label{appb}
In this appendix, we discuss gauging subgroup of internal symmetries in a system with LSM anomaly that we introduced in the main text gives modulated symmetry -- consistent with the previous study~\cite{Ebisu:2025mtb}. \par
We recall that the Hamiltonian is described by
\begin{eqnarray}
    H_{2D}=&-&J_x\sum_{\br}Z_{\br-\ex}Z_{\br+\ex}-J_y\sum_{\bp}Z_{\bp-\ey}Z_{\bp+\ey}\nonumber\\
      &-&J_G\sum_{\br}X_{\br-\ex}X_{\br+\ex}X_{\br-\ey}X_{\br+\ey}-J_B\sum_\bp Z_{\bp-\ex}Z_{\bp+\ex}Z_{\bp+\ey}Z_{\bp-\ey}\label{b1}
\end{eqnarray}
with $J_B\to\infty$. The model is on square lattice with periodic boundary condition with system size~$L_x\times L_y$, where we set $L_x$ to be even. 
This model respects $0$-form and $1$-form global symmetries~\eqref{symmetry}. In the following, we gauge one of the global symmetries to see emergence of the modulated symmetries. 
\subsection{Gauging $0$-form symmetry}
We first focus on gauging $\mathbb{Z}_2^{(0)}$. The Gauss's law reads
\begin{eqnarray}
   X_{\br+\ex}\widehat{\tau}^X_{\br+\exx}\widehat{\tau}^X_{\br}\widehat{\mu}^X_{\br+\ex+\ey}\widehat{\mu}^X_{\br+\ex-\ey}=1, \label{gauss3}
\end{eqnarray}
where $\widehat{\tau}^X_{\br}$ and $\widehat{\mu}^X_{\bp}$ denote Pauli $X$ operators that act on extended Hilbert space on each node and plaquette.  
Accordingly, spin~$Z$ terms in the Hamiltonian~\eqref{b1} are modified so that they commute with the Gauss's law~\eqref{gauss3}:
\begin{eqnarray}
    Z_{\br-\ex}Z_{\br+\ex}\to Z_{\br-\ex}\widehat{\tau}^Z_\br Z_{\br+\ex},\quad Z_{\bp-\ey}Z_{\bp+\ey}\to Z_{\bp-\ey}\widehat{\mu}^Z_\bp Z_{\bp+\ey}\nonumber\\
    Z_{\bp-\ex}Z_{\bp+\ex}Z_{\bp+\ey}Z_{\bp-\ey}\to Z_{\bp-\ex}Z_{\bp+\ex}Z_{\bp+\ey}Z_{\bp-\ey}\widehat{\mu}^Z_{\bp}. \label{three}
\end{eqnarray} 
We rewrite variables as
\begin{eqnarray}
    \tau^X_\br\vcentcolon&=&\widehat{\tau}^X_\br,\quad \tau^Z_\br\vcentcolon=Z_{\br-\ex}\widehat{\tau}^Z_\br Z_{\br+\ex}\nonumber\\
    \mu^X_\bp\vcentcolon&=&\widehat{\mu}^X_\bp,\quad \mu^Z_\bp\vcentcolon=
    Z_{\bp+\ey}Z_{\bp-\ey}\widehat{\mu}^Z_{\bp}\label{kk}
\end{eqnarray}
Further, we add the following gauge flux 
\begin{eqnarray}
    -J_M\sum_\br \tau^Z_\br \tau_{\br+\eyy}^Z Z_{\br+\exx+\ey}Z_{\br-\exx+\ey}
\end{eqnarray}
with $J_M\to\infty$
to the Hamiltonian~\eqref{b1} to make the theory fluxless. 
Since $J_B\to\infty$ in the original Hamiltonian~\eqref{b1}, and \eqref{three}\eqref{kk}, we have the following constraint:
\begin{eqnarray}
    Z_{\bp-\ex}\mu^Z_\bp Z_{\bp+\ex}=1\quad\forall \bp.
\end{eqnarray}
From this constraint, jointly with substituting the Gauss law~\eqref{gauss3} to the third term of the Hamiltonian~\eqref{b1} and decoupling $\mu^X_{\bp}$ from the theory, we arrive at the following gauged Hamiltonian:
\begin{eqnarray}
    H_{2D/\mathbb{Z}^{(0)}_2}=&-&J_x\sum_{\br}\tau^Z_\br-J_y\sum_\br Z_{\br+\ey}Z_{\br+\exx+\ey}\nonumber\\
    &-&J_G\sum_\br \tau^X_{\br-\exx}\tau^X_{\br+\exx}X_{\br-\ey}X_{\br+\ey}-J_M\sum_{\br}\tau^Z_\br \tau^Z_{\br+\eyy}Z_{\br-\exx+\ey}Z_{\br+\exx+\ey}\label{dipoletoric}
\end{eqnarray}
with $J_M\to\infty$.
The last two terms in~\eqref{dipoletoric} constitute the toric code-like stabilizer model with dipole~$1$-form symmetries that was studied in~\cite{10.21468/SciPostPhys.14.5.106,Gorantla:2022pii}. Indeed, the model~\eqref{dipoletoric} admits the following~$1$-form symmetries, corresponding to noncontractible loops in the $x$- and $y$-direction:
\begin{align}  
    \xi^Z_0&:=\prod_{\hx=1}^{L_x}\tau^Z_{(\hx,1)},\quad \xi^Z_{x}\vcentcolon=\prod_{\hx=1}^{L_x}\left(\tau^Z_{(\hx,1)}\right)^{\hx}\nonumber\\
    \zeta^Z_0&:=\prod_{\hy=1}^{L_y}Z_{(1,\frac{1}{2}+\hy)}Z_{(2,\frac{1}{2}+\hy)},\quad \zeta^Z_x\vcentcolon=\prod_{\hy=1}^{L_y}Z_{(1,\frac{1}{2}+\hy)}.
\end{align}
One can verify that 
\begin{eqnarray}
    T_x\xi^Z_{x}T_x^{-1}=\xi^Z_0\xi^Z_x\nonumber\\
    T_x\zeta_x^ZT_x^{-1}=\zeta_0^Z\zeta_x^Z\label{1fomrd}
\end{eqnarray}
with other combinations between $1$-form symmetries and translation operators, including the one in $y$-direction, give trivial commutation relation. Relations~\eqref{1fomrd} imply that the 1-form symmetries are subject to \emph{1-form dipole algebra}~\cite{Ebisu:2024eew}, that is $1$-form analog of~\eqref{dipole0}.
\subsection{Gauging $1$-form symmetry}
Now we turn to gauging $1$-form symmetry in the model~\eqref{b1}. To do so, we introduce extended Hilbert space on each node of the lattice whose Pauli operator is denoted as $\sigma^{X/Z}_\br$ and impose the following Gauss's law:
\begin{eqnarray}
    \sigma^Z_\br Z_{\br+\ex}\sigma^Z_{\br+\exx}=1,\quad \sigma^Z_\br Z_{\br+\ey}\sigma^Z_{\br+\eyy}=1.
\end{eqnarray}
Accordingly, the third term in~\eqref{b1} is modified as 
\begin{eqnarray}
    X_{\br-\ex}X_{\br+\ex}X_{\br-\ey}X_{\br+\ey}\to X_{\br-\ex}X_{\br+\ex}\sigma^X_\br X_{\br-\ey}X_{\br+\ey}.
\end{eqnarray}
The gauged Hamiltonian reads
\begin{eqnarray}
    H_{2D/\mathbb{Z}_2^{(1)}}=&-&J_x\sum_{\br}\sigma^Z_{\br-\exx}\sigma^Z_{\br+\exx}-J_y\sum_{\br}\sigma^Z_{\br}\sigma^Z_{\br+\exx}\sigma^Z_{\br+\exx+\eyy}\sigma^Z_{\br+\eyy}\nonumber\\&-&
      J_G\sum_{\br}X_{\br-\ex}X_{\br+\ex}\sigma^X_{\br}X_{\br-\ey}X_{\br+\ey}.
\end{eqnarray}
The gauged Hamiltonian respects the following $0$-form symmetries:
\begin{eqnarray}
    Q_{2D,0}\vcentcolon=\prod_{\hx=1}^{L_x}\prod_{\hy=1}^{L_y}\sigma^X_\br,\quad  Q_{2D,x}\vcentcolon=\prod_{\hx=1}^{L_x}\prod_{\hy=1}^{L_y}\left(\sigma^X_{\br}\right)^{\hx}.
\end{eqnarray}
These two symmetries are subject to
\begin{eqnarray}
 T_x Q_{2D,0}T_x^{-1}= Q_{2D,0},\quad 
    T_x Q_{2D,x}T_x^{-1}=Q_{2D,0}Q_{2D,x},
\end{eqnarray}
indicating that these two charges constitute $0$-form dipole algebra. \par
In summary, we implement gauging subgroup of internal symmetries in a system with LSM anomaly~\eqref{lsm} to see modulated symmetry emerges. More explicitly, gauging $0$-form ($1$-form) symmetry leads to $1$-form ($0$-form) modulated symmetry. 
\section{Some calculations regarding duality mapping}\label{app22}
In this appendix, we give derivations of~\eqref{56} and the fusion rule~\eqref{fusion2} of the lattice translation operators. 
\subsection{Derivation of~\eqref{56}}
From~\eqref{phasefactor}, we have
\begin{eqnarray}
  D\sigma^X_{\br_0} &=& \frac{1}{\sqrt{2}}  \sum_{\substack{\{\tau_{\br}\},\{\sigma_{\br}\},\{\mu_{\bp}\} \\ \{\tau^\prime_{\br}\},\{\sigma^\prime_{\br}\},\{\mu^\prime_{\bp}\} }}\Omega\ket{\{\tau^\prime_\br\},\{\sigma^\prime_\br\},\{\mu^\prime_\bp\}}\bra{\{\tau_\br\},\{\sigma_\br\},\{\mu_\bp\}}\sigma^X_{\br_0}\nonumber\\
  &=& \frac{1}{\sqrt{2}} \sum_{\substack{\{\tau_{\br}\},\{\sigma_{\br}\},\{\mu_{\bp}\} \\ \{\tau^\prime_{\br}\},\{\sigma^\prime_{\br}\},\{\mu^\prime_{\bp}\} }}\Omega\ket{\{\tau^\prime_\br\},\{\sigma^\prime_\br\},\{\mu^\prime_\bp\}}\bra{\{\tau_\br\},\{\sigma_{\br\neq\br_0};\sigma_{\br_0}+1\},\{\mu_\bp\}}\nonumber\\
  &=&\frac{1}{\sqrt{2}}  \sum_{\substack{\{\tau_{\br}\},\{\sigma_{\br}\},\{\mu_{\bp}\} \\ \{\tau^\prime_{\br}\},\{\sigma^\prime_{\br}\},\{\mu^\prime_{\bp}\} }}(-1)^{\tau^\prime_{\br_0+\exx}+\tau^\prime_{\br_0}+\mu^\prime_{\br_0+\ex+\ey}+\mu^\prime_{\br_0+\ex-\ey}}\times\Omega\ket{\{\tau^\prime_{\br_0}\},\{\sigma^\prime_\br\},\{\mu^\prime_\bp\}}\bra{\{\tau_\br\},\{\sigma_\br\},\{\mu_\bp\}}\nonumber\\
  &=&\tau^Z_{\br_0+\exx}\tau^Z_{\br_0}\mu^Z_{\br_0+\ex+\ey}\mu^Z_{\br_0+\ex-\ey}\frac{1}{\sqrt{2}} \sum_{\substack{\{\tau_{\br}\},\{\sigma_{\br}\},\{\mu_{\bp}\} \\ \{\tau^\prime_{\br}\},\{\sigma^\prime_{\br}\},\{\mu^\prime_{\bp}\} }}\Omega\ket{\{\tau^\prime_\br\},\{\sigma^\prime_\br\},\{\mu^\prime_\bp\}}\bra{\{\tau_\br\},\{\sigma_\br\},\{\mu_\bp\}}\nonumber\\
  &=&\tau^Z_{\br_0+\exx}\tau^Z_{\br_0}\mu^Z_{\br_0+\ex+\ey}\mu^Z_{\br_0+\ex-\ey}D, \label{99}
\end{eqnarray}
giving the first relation in~\eqref{56}. 
From the second to third line of~\eqref{99}, we relabel the variable as~$\tau_{\br_0}+1\to\tau_{\br_0}$. The other relations in~\eqref{56} are similarly derived. 
\subsection{Derivation of the fusion rule}
As for the derivation of fusion rule~\eqref{fusion2}, we first evaluate $D\times D$, which is written as
\begin{eqnarray}
    D\times D&=&\frac{1}{2} \sum_{\substack{\{\alpha_{\br}\},\{\beta_{\br}\},\{\gamma_{\bp}\} \\ \{\alpha^\prime_{\br}\},\{\beta^\prime_{\br}\},\{\gamma^\prime_{\bp}\} }} \sum_{\substack{\{\tau_{\br}\},\{\sigma_{\br}\},\{\mu_{\bp}\} \\ \{\tau^\prime_{\br}\},\{\sigma^\prime_{\br}\},\{\mu^\prime_{\bp}\} }}\Xi\times\Omega\nonumber\\
   & \times&\ket{\{\alpha^\prime_\br\},\{\beta^\prime_\br\},\{\gamma^\prime_\bp\}}\langle{\{\alpha_\br\},\{\beta_\br\},\{\gamma_\bp\}}
    \ket{\{\tau^\prime_\br\},\{\sigma^\prime_\br\},\{\mu^\prime_\bp\}}\bra{\{\tau_\br\},\{\sigma_\br\},\{\mu_\bp\}}.\label{alpha}
\end{eqnarray}
Here, 
\begin{eqnarray}
    \Xi=(-1)^{\sum_{\br}\left(\alpha^\prime_{\br+\exx}+\alpha^\prime_{\br}+\gamma^\prime_{\br+\ex+\ey}+\gamma^\prime_{\br+\ex-\ey}\right)\beta_{\br}+(\beta^\prime_{\br-\exx}+\beta^\prime_{\br})\alpha_\br+(\beta^\prime_{\br+\eyy}+\beta^\prime_{\br})\gamma_{\br+\ex+\ey}},
\end{eqnarray}
which is obtained from~\eqref{phasefactor} by replacing indices as $\tau_\br\to \alpha_\br$, $\sigma_\br\to\beta_\br$, $\mu_\bp\to\gamma_\bp$ and similarly for indices with prime.
Since
\begin{eqnarray}
    \langle{\{\alpha_\br\},\{\beta_\br\},\{\gamma_\bp\}}
    \ket{\{\tau^\prime_\br\},\{\sigma^\prime_\br\},\{\mu^\prime_\bp\}}=\prod_{\br}\left[\delta_{\alpha_\br,\tau^\prime_\br}\times \delta_{\beta_\br,\sigma^\prime_\br}\right]\times\prod_{\bp}\delta_{\gamma_\bp,\mu^\prime_\bp},
\end{eqnarray}
the right hand side of~\eqref{alpha} is simplified as
\begin{eqnarray}
 \frac{1}{2}  \sum_{\{\tau^\prime_\br\},\{\sigma^\prime_\br\},\{\mu^\prime_\bp\}} \sum_{\substack{\{\tau_{\br}\},\{\sigma_{\br}\},\{\mu_{\bp}\} \\ \{\alpha^\prime_{\br}\},\{\beta^\prime_{\br}\},\{\gamma^\prime_{\bp}\} }} \Theta \ket{\{\alpha^\prime_\br\},\{\beta^\prime_\br\},\{\gamma^\prime_\bp\}}\bra{\{\tau_\br\},\{\sigma_\br\},\{\mu_\bp\}}\label{hi}
\end{eqnarray}
with 
\begin{eqnarray}
    \Theta&=&\exp\left[i\pi \sum_{\br}\left(\alpha^\prime_{\br+\exx}+\alpha^\prime_{\br}+\gamma^\prime_{\br+\ex+\ey}+\gamma^\prime_{\br+\ex-\ey}+\tau_{\br+\exx}+\tau_{\br}+\mu_{\br+\ex+\ey}+\mu_{\br+\ex-\ey}\right)\sigma^\prime_{\br}\right]\nonumber\\
    &\times&\exp\left[i\pi\sum_{\br}(\beta^\prime_{\br-\exx}+\beta^\prime_{\br}+\sigma_{\br-\exx}+\sigma_\br)\tau^\prime_\br
    +(\beta^\prime_{\br+\eyy}+\beta^\prime_{\br}+\sigma_{\br+\eyy}+\sigma_\br)\mu^\prime_{\br+\ex+\ey}
    \right].
      \end{eqnarray}
Summing over variables $\{\tau^\prime_\br\},\{\sigma^\prime_\br\},\{\mu^\prime_\bp\}$ in~\eqref{hi} gives the following constraints:
\begin{eqnarray}
   \alpha^\prime_\br&=& \tau_\br +k_\br,\quad \alpha^\prime_{\br+\exx}= \tau_{\br+\exx} +k_{\br+\exx},\nonumber\\
   \gamma^\prime_{\br+\ex+\ey}&=&  \mu^\prime_{\br+\ex+\ey}+l_{\br+\ex+\ey},\quad
   \gamma^\prime_{\br+\ex-\ey}=\mu_{\br+\ex-\ey} +l_{\br+\ex-\ey}\nonumber\\
   (k_{\br},k_{\br+\exx},l_{\br+\ex+\ey},l_{\br+\ex-\ey}&\in&\mathbb{Z}_2, \quad
   k_{\br}+k_{\br+\exx}+l_{\br+\ex+\ey}+l_{\br+\ex-\ey}=0\mod 2)\label{star}
\end{eqnarray}
\begin{eqnarray}
    \beta^\prime_{\br}=\sigma_\br+s,\quad (s\in\mathbb{Z}_2).\label{flip}
\end{eqnarray}
Similar to the ground state of the toric code, the constraint in~\eqref{star} enforces a local parity condition that permits only an even number of spin flips at each site~$\br$. Furthermore, since we focus on the projected state [i.e., the ground state with a trivial eigenvalue of the flux operator in~\eqref{flux2}], there are four distinct configurations of spin flips corresponding to the non-trivial homology classes of the torus. One such an example is given by
\begin{eqnarray}
    k_{\br}&=&\begin{cases}
        k_{(\hx,1)}=1\quad (1\leq \hx\leq L_x)\\
        0\quad (\text{else})
    \end{cases},\nonumber\\
    l_{\br+\ex+\ey}&=&0\quad\forall \br,
\end{eqnarray}
that is, trajectory of the spin flip forms noncontractible loop in the $x$-direction. \par
Substituting constraints~\eqref{star}~\eqref{flip} into~\eqref{hi}, we have
\begin{eqnarray}
    D\times D&=&  \nonumber\\
  \frac{1}{2}\sum_{a,b,c=0,1} \sum_{\{\tau_\br\},\{\sigma_\br\},\{\mu_\bp\}}&&\left(\prod_{\br}\sigma^X_{\br}\right)^a\left(\prod_{\hx=1}^{L_x}\tau^X_{(\hx,1)}\right)^b \left(\prod_{\hy=1}^{L_y}\mu^X_{(\frac{1}{2},\frac{1}{2}+\hy)}\right)^c\ket{\{\tau_\br\},\{\sigma_\br\},\{\mu_\bp\}}\bra{\{\tau_\br\},\{\sigma_\br\},\{\mu_\bp\}}\nonumber,
\end{eqnarray}
from which, jointly with the fact that $(CZ)^2=1$, where $CZ$ is product of control phase gate operators~\eqref{35}, we finally arrive at the fusion rule~\eqref{fusion2} in the main text.

\section{3D lattice model}\label{3da}
\begin{figure}
    \begin{center}
         \begin{subfigure}[h]{0.59\textwidth}
       \centering
  \includegraphics[width=0.9\textwidth]{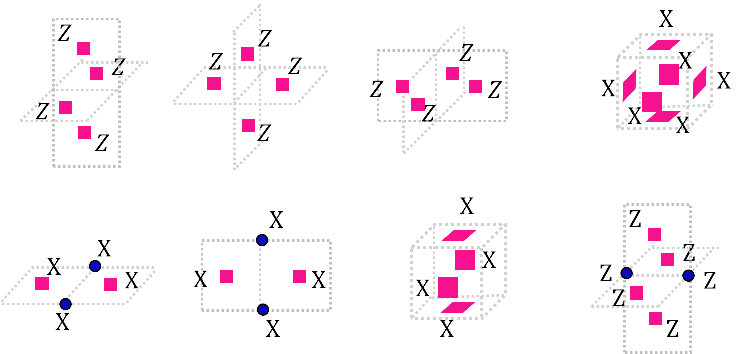}
         \caption{}\label{3d1}
             \end{subfigure}
            \begin{subfigure}[h]{0.40\textwidth}
            \centering
  \includegraphics[width=1.0\textwidth]{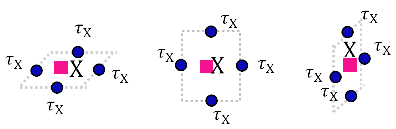}
         \caption{}\label{gauss3d}
             \end{subfigure}
                         \begin{subfigure}[h]{0.68\textwidth}
            \centering
  \includegraphics[width=1.1\textwidth]{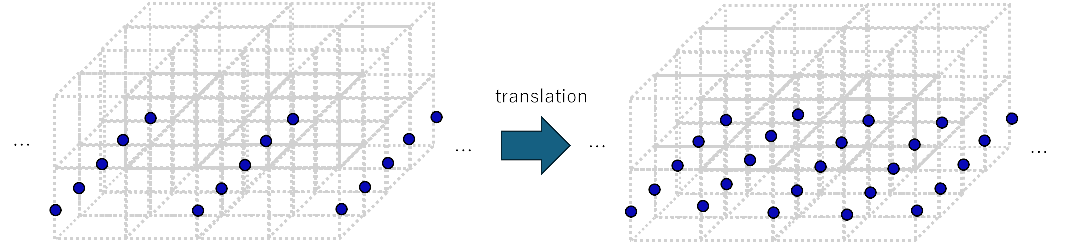}
         \caption{}\label{dipole3d}
             \end{subfigure}
          \end{center}
 \caption{(a) Configurations of terms that constitute the Hamiltonian~\eqref{hami3}. The first three configurations correspond to the first term in~\eqref{hami3}, whereas the fourth, fifth, sixth, seventh, eighth configurations do to the second, third, fourth, fifth, and sixth term in~\eqref{hami3}, respectively. (b) The Gauss's law when gauging one of the $1$-form symmetries~\eqref{mem1}, described by~\eqref{gauss3d1}. (c)~Illustration for one of examples of the $1$-form dipole algebra~\eqref{103} with $ab=xy$; acting a translation operator $T_x$ on $\mathcal{M}^{Z(1)}_{xy,x}$ (left) gives $\mathcal{M}^{Z(1)}_{xy,0}$ (right).    }
 \end{figure}
In this Appendix, we study a 3D lattice model with the LSM anomaly, that is, mixed anomaly between two $1$-form and translation symmetries. In this model, we first investigate emergence of the modulated symmetries via gauging one of $1$-form symmetries. Next, we turn to gauging both $1$-form symmetries to see that gauge invariant lattice translation operator becomes non-invertible. These line of arguments are consistent with the diagram of Fig.~\ref{diagram}, viz, in the system with LSM anomaly, gauging subgroup of internal symmetries gives rise to modulated symmetries whereas gauging full internal symmetries yields non-invertible translation. Similar to consideration in 2D (Sec.~\ref{sec3}, Appendix.~\ref{appb}), we focus on the case with $N=2$.
\subsection{Model and symmetries}
We envisage a cubic model where a spin-$1/2$ d.o.f on each plaquette and node whose Pauli $Z$ operator is denoted as $Z_{\bp_{ab}}$~$(ab=xy,yz,zx)$ and $Z_\br$ (similarly for Pauli $X$ operator). Here, $\bp_{ab}$ represents coordinate of a plaquette on $xy$-, $yz$-, and $zx$-plane. More explicitly, 
\begin{equation*}
    \bp_{xy}\vcentcolon=(\hx+\frac{1}{2},\hy+\frac{1}{2},\hz),\quad  \bp_{yz}\vcentcolon=(\hx,\hy+\frac{1}{2},\hz+\frac{1}{2}),\quad \bp_{zx}\vcentcolon=(\hx+\frac{1}{2},\hy,\hz+\frac{1}{2}).
\end{equation*}
Also, for latter convenience, we introduce coordinate of links and center of a cube as 
\begin{eqnarray*}
    \bx\vcentcolon= (\hx+\frac{1}{2},\hy,\hz),\quad  \by\vcentcolon= (\hx,\hy+\frac{1}{2},\hz),\quad  \bz\vcentcolon= (\hx,\hy,\hz+\frac{1}{2}),\quad\bc\vcentcolon=(\hx+\frac{1}{2},\hy+\frac{1}{2},\hz+\frac{1}{2}).
\end{eqnarray*}
The first three coordinates are abbreviated as $\bl_a~(a=x,y,z)$.
The Hamiltonian is given by
\begin{eqnarray}
    H_{3D}=&-&J_1\sum_{\bl_{a}}\prod_{ \bp_{a^\prime b^\prime}\supset\bl_{a}}Z_{\bp_{a^\prime b^\prime}}-J_2\sum_{\bc}\prod_{\bp_{ab}\subset \bc}X_{\bp_{ab}}\nonumber\\
    &-&J_3\sum_{\bl_y}\left(\prod_{\bp_{xy}\supset \bl_y}X_{\bp_{xy}}\times\prod_{\br\subset \bl_y}X_\br\right)  -J_4\sum_{\bl_z}\left(\prod_{\bp_{zx}\supset \bl_z}X_{\bp_{zx}}\times\prod_{\br\subset \bl_z}X_\br\right)
    \nonumber\\
    &-&J_5\sum_\bc\left(\prod_{\bp_{xy}\subset \bc}X_{\bp_{xy}}\times\prod_{\bp_{zx}\subset \bc}X_{\bp_{zx}}\right)
    -J_6\sum_{\bl_x}\left(\prod_{\bp_{ab}\supset \bl_x}Z_{\bp_{ab}}\times\prod_{\br\subset \bl_x}Z_{\br}\right),\label{hami3}
\end{eqnarray}
where the product ``$\prod_{ \bp_{a^\prime b^\prime}\supset\bl_a}$'' in the first term means that 
operators $Z_{\bp_{a^\prime b^\prime}}$ on plaquettes that have overlap with a given link $\bl_a$ are multiplied (Other products are analogously understood. ) See Fig.~\ref{3d1} for visual illustration. We further assume that $J_2\to\infty$ and $J_6\to \infty$ so that we focus on a project state with trivial eigenvalue of the second and last terms of~\eqref{hami3}. Note that the first two terms in~\eqref{hami3} correspond to the 3D toric code. \par
The model has two $1$-form symmetries, $\mathbb{Z}_2^{(1)}\times \mathbb{Z}_2^{(1)}$, described by membrane operators. One of the $1$-form symmetries comes from the the 3D toric code, described by
\begin{eqnarray}
    U^{X(1)}_{xy}\vcentcolon&=&\prod_{\hx=1}^{L_x}\prod_{\hy=1}^{L_y}X_{(\hx+\frac{1}{2},\hy+\frac{1}{2},1)},\nonumber\\
    U^{X(1)}_{yz}\vcentcolon&=&\prod_{\hy=1}^{L_y}\prod_{\hz=1}^{L_z}X_{(1,\hy+\frac{1}{2},\hz+\frac{1}{2})},\nonumber\\
    U^{X(1)}_{zx}\vcentcolon&=&\prod_{\hz=1}^{L_z}\prod_{\hx=1}^{L_x}X_{(\hx+\frac{1}{2},1,\hz+\frac{1}{2})}.\label{mem1}
    \end{eqnarray}
Another $1$-form symmetry is described by the following membrane operators:
\begin{eqnarray}
    V^{Z(1)}_{xy}\vcentcolon=\prod_{\hx=1}^{L_x}\prod_{\hy=1}^{L_y}Z_{(\hx+\frac{1}{2},\hy,\frac{3}{2})},\quad V^{Z(1)}_{yz}\vcentcolon=\prod_{\hy=1}^{L_y}\prod_{\hz=1}^{L_z}Z_{(1,\hy,\hz)},\quad V_{zx}^{Z(1)}\vcentcolon=\prod_{\hz=1}^{L_z}\prod_{\hx=1}^{L_x}Z_{(\hx+\frac{1}{2},\frac{3}{2},\hz)}\label{mem2}
\end{eqnarray}
  Note that these membrane operators~\eqref{mem1}\eqref{mem2} are topological, i.e., independent operators depend solely on the homology class due to the assumption that we take the limit $J_2,J_6\to \infty$. The two $1$-form symmetry operators exhibit nontrivial commutation relation, depending on the system size in the $x$-direction:
  \begin{eqnarray}
      U_{xy}^{X(1)}V_{zx}^{Z(1)}=(-1)^{L_x}V_{zx}^{Z(1)}U_{xy}^{X(1)},\quad U_{zx}^{X(1)}V_{xy}^{Z(1)}=(-1)^{L_x}V_{xy}^{Z(1)}U^{X(1)}_{zx},\label{103}
  \end{eqnarray}
which signals the LSM anomaly. In the next subsection, we gauge one of the $1$-form symmetries,~\eqref{mem1}.
\subsection{Gauging one of $1$-form symmetries }
In what follows, we set $L_x$ to be even. 
To gauge $1$-form symmetry~\eqref{mem1}, we accommodate extended Hilbert space on link whose Pauli operator is denoted as $\tau^{X/Z}_{\bl_a}$. The Gauss's law reads
\begin{eqnarray}
    X_{\bp_{ab}}\times \prod_{\bl_{a^\prime}\subset\bp_{ab}} \tau^X_{\bl_{a^\prime}}=1,\quad\forall\bp_{ab}.\label{gauss3d1}
\end{eqnarray}
See also Fig.~\ref{gauss3d}. Accordingly, the first and last term in~\eqref{hami3} is modified as
\begin{eqnarray}
    \prod_{ \bp_{a^\prime b^\prime}\supset\bl_{a}}Z_{\bp_{a^\prime b^\prime}}&\to& \prod_{ \bp_{a^\prime b^\prime}\supset\bl_{a}}Z_{\bp_{a^\prime b^\prime}}\times\tau^Z_{\bl_a},\quad\forall \bl_a\nonumber\\
    \prod_{\bp_{ab}\supset \bl_x}Z_{\bp_{ab}}\times\prod_{\br\subset \bl_x}Z_{\br}&\to& \prod_{\bp_{ab}\supset \bl_x}Z_{\bp_{ab}}\times\prod_{\br\subset \bl_x}Z_{\br}\times \tau^X_{\bl_x}\quad\forall \bl_x,
\end{eqnarray}
Physically, this procedure is the minimal coupling to the gauge fields. Further, we add the following operator to make the theory so that it does not admit excess gauge flux:
\begin{eqnarray}
    -J_7\sum_\br\prod_{\bl_a\supset\br}\tau^X_{\bl_a}.
\end{eqnarray}
Substituting the Gauss's law into the other terms in~\eqref{hami3}, we have the following gauged Hamiltonian:
\begin{eqnarray}
    H_{3D/\mathbb{Z}_2^{(1)}}=&-&J_1\sum_{\bl_a}\prod_{ \bp_{a^\prime b^\prime}\supset\bl_{a}}Z_{\bp_{a^\prime b^\prime}}\times\tau^Z_{\bl_a}-J_3\sum_{\bl_y}\left(\prod_{s,t,u=\pm1}\tau^X_{\bl_y+s\exx}\tau^X_{\bl_y-\ey+t\ex}\tau^X_{\bl_y+\ey+u\ex}\times\prod_{\br\subset\bl_y}X_{\br}\right)\nonumber\\
    &-&J_4\sum_{\bl_z}\left(\prod_{s,t,u=\pm1}\tau^X_{\bl_z+s\exx}\tau^X_{\bl_z-\ez+t\ex}\tau^X_{\bl_z+\ez+u\ex}\times\prod_{\br\subset\bl_z}X_{\br}\right)-J_5\sum_{\bc}\left(\prod_{\bl_y\subset\bc}\tau^X_{\bl_y}\times\prod_{\bl_z\subset\bc}\tau^X_{\bl_z}\right)\nonumber\\
   & -&J_6\sum_{\bl_x}\left( \prod_{\bp_{ab}\supset \bl_x}Z_{\bp_{ab}}\times\prod_{\br\subset \bl_x}Z_{\br}\times \tau^X_{\bl_x}\right) -J_7\sum_\br\prod_{\bl_a\supset\br}\tau^X_{\bl_a}\label{gauged3d}
\end{eqnarray}
with $J_6,J_7\to\infty$~\footnote{The second term in the original Hamiltonian~\eqref{hami3} becomes identity after gauging $1$-form symmetry. }. 
The gauged Hamiltonian~\eqref{gauged3d} respects the following dual $1$-form symmetries:
\begin{eqnarray}
    \mathcal{M}^{Z(1)}_{xy,x}\vcentcolon=\prod_{\hx=1}^{L_x}\prod_{\hy=1}^{L_y}\left(\tau^Z_{(\hx,\hy,\frac{3}{2})}\right)^{\hx},\quad M^{Z(1)}_{xy,0}\vcentcolon=\prod_{\hx=1}^{L_x}\prod_{\hy=1}^{L_y}\tau^Z_{(\hx,\hy,\frac{3}{2})}\nonumber\\
        \mathcal{M}^{Z(1)}_{yz,x}\vcentcolon=\prod_{\hy=1}^{L_y}\prod_{\hz=1}^{L_z}Z_{(1,\hy,\hz)},\quad \mathcal{M}^{Z(1)}_{yz,0}\vcentcolon=\prod_{\hy=1}^{L_y}\prod_{\hz=1}^{L_z}\tau^Z_{(\frac{3}{2},\hy,\hz)}\nonumber\\
           \mathcal{M}^{Z(1)}_{zx,x}\vcentcolon=\prod_{\hz=1}^{L_z}\prod_{\hx=1}^{L_x}\left(\tau^Z_{(\hx,\frac{3}{2},\hz)}\right)^{\hx},\quad \mathcal{M}^{Z(1)}_{zx,0}\vcentcolon=\prod_{\hx=1}^{L_x}\prod_{\hz=1}^{L_z}\tau^Z_{(\hx,\frac{3}{2},\hz)}.
\end{eqnarray}
Furthermore, these symmetries satisfy the following relations:
\begin{eqnarray}
    T_x\mathcal{M}^{Z(1)}_{ab,x}T_x^{-1}=\mathcal{M}^{Z(1)}_{ab,x}\mathcal{M}^{Z(1)}_{ab,0},\quad (ab=xy,yz,zx),\label{abxy}  
\end{eqnarray}
with other relations involving charges and translation operators, including the ones in the $y$- and $z$-direction being trivial. 
The relation in~\eqref{abxy} with $ab=xy$ is portrayed in Fig.~\ref{dipole3d}.
Hence, we obtain $1$-form dipole symmetry. The same conclusion is drawn, i.e., the $1$-form modulated symmetry has emerged, when we perform gauging the other $1$-form symmetry~\eqref{mem2} instead of the one in~\eqref{mem1}. 

\subsection{Gauging full internal symmetries}
We move on to gauging both $1$-form symmetries~\eqref{mem1}\eqref{mem2} to see how non-invertible translation operator emerges. Since the discussion closely parallels the one in the case of 2D (Sec.~\ref{sec3}), we demonstrate construction of the non-invertible translation succinctly. \par
To gauge both of symmetries~\eqref{mem1}\eqref{mem2}, we introduce an extended Hilbert space on 
each $y$- and $z$-link, another one on each link, the other on each center of a cube with corresponding Pauli operators being denoted by $\sigma^{X/Z}_{\bl_a}~(a=y,z)$, $\tau^{X/Z}_{\bl_a}~(a=x,y,z)$, and $\mu^{X/Z}_{\bc}$, respectively. We 
impose the following Gauss's laws:
\begin{eqnarray}
    1&=&Z_{\bp_{xy}}\times\left(\prod_{\bl_y\subset\bp_{xy}}\sigma^Z_{\bl_y}\prod_{\bc\supset\bp_{xy}}\mu^Z_{\bc}\right)\times \left(\tau^X_{\bp_{xy}+\ex}\right)^{\hx+1}
    =
    \vcenter{\hbox{
        \begin{tikzpicture}[tdplot_main_coords]
            \draw[dashed](0,0,0)--++(1,0,0)--++(0,1,0)--++(-1,0,0)--cycle;
            \fill[red](.325,.325,0)--++(.25,0,0)--++(0,.25,0)--++(-.25,0,0)--cycle;
            \node at(.675,.675,0)[red]{$Z$};
            \fill[green](0,.5,0)circle(.1)node[left]{$\sigma^Z$};
            \fill[green](1,.5,0)circle(.1);
            \fill[blue](.5,.5,.5)circle(.1)node[above]{$\mu^Z$};
            \fill[blue](.5,.5,-.5)circle(.1)node[below]{$\mu^Z$};
            \draw(1,.5,0)circle(.15)node[right]{$(\tau^X)^{\hat{x}+1}$};
        \end{tikzpicture}
    }}(=\vcentcolon G^Z_{\bp_{xy}})
    \quad\forall \bp_{xy}\nonumber\\
1&=&Z_\br\times\left(
\prod_{\substack{\bl_a\supset\br \\ a=y,z}}\sigma^Z_{\bl_a}\right)\times \left(\tau^X_{\br+\ex}\right)^{\hx}
=\vcenter{\hbox{
    \begin{tikzpicture}[tdplot_main_coords]
        \draw[dashed](-1,0,0)--(1,0,0);
        \draw[dashed](0,-1,0)--(0,1,0);
        \draw[dashed](0,0,-1)--(0,0,1);
        \fill[blue](0,0,0)circle(.1)node[above left]{$Z$};
        \foreach\i in{1,-1}{
            \fill[green](0,\i/2,0)circle(.1);
            \fill[green](0,0,\i/2)circle(.1);
        }
        \fill(.5,0,0)circle(.1)node[below right]{$(\tau^X)\hat{x}$};
        \node at(0,0,-.5)[below left,green]{$\sigma^Z$};
    \end{tikzpicture}
}}
(=\vcentcolon G^Z_\br)
\quad\forall\br\nonumber\\
1&=&Z_{p_{zx}}\times\left(
\prod_{\bl_z\subset\bp_{zx}}\sigma^Z_{\bl_z}\prod_{\bc\supset\bp_{zx}}\mu^Z_{\bc}\right)\times \left(\tau^X_{\bp_{zx}+\ex}\right)^{\hx+1}
=
\vcenter{\hbox{
    \begin{tikzpicture}[tdplot_main_coords]
        \draw[dashed](0,0,0)--++(1,0,0)--++(0,0,1)--++(-1,0,0)--cycle;
        \fill[red](.325,0,.325)--++(.25,0,0)--++(0,0,.25)--++(-.25,0,0)--cycle;
        \fill[blue](.5,.5,.5)circle(.1)node[above right]{$\mu^Z$};
        \fill[blue](.5,-.5,.5)circle(.1)node[below left]{$\mu^Z$};
        \foreach\i in{0,1}{
            \fill[green](\i,0,.5)circle(.1);
        }
        \draw(1,0,.5,)circle(.15)node[right]{$(\tau^X)^{\hat{x}+1}$};
        \node at(0,0,.5)[left,green]{$\sigma^Z$};
        \node at(.325,0,.625)[red]{$Z$};
    \end{tikzpicture}
}}
(=\vcentcolon G^Z_{\bp_{zx}})\quad\forall\bp_{zx}\nonumber\\
1&=&X_{\bp_{xy}}\times \left(\prod_{\bl_a\subset \bp_{xy}}\tau^Z_{\bl_a}\right)\times \left(\sigma^X_{\bp_{xy}-\ex}\right)^{\hx}
=
\vcenter{\hbox{
            \begin{tikzpicture}[tdplot_main_coords]
                \draw[dashed](0,0,0)--++(1,0,0)--++(0,1,0)--++(-1,0,0)--cycle;
                \fill[red](.325,.325,0)--++(.25,0,0)--++(0,.25,0)--++(-.25,0,0)--cycle;
                \node at(.625,.325,0)[red,above]{$X$};
                \foreach\i in{0,1}{
                    \fill(\i,.5,0)circle(.1);
                    \fill(.5,\i,0)circle(.1);
                }
                \node at(1,.5,0)[right]{$\tau^Z$};
                \draw[green](0,.5,0)circle(.15)node[left]{$(\sigma^X)^{\hat{x}}$};
            \end{tikzpicture}
        }}
        (=\vcentcolon G^X_{\bp_{xy}})\quad\forall\bp_{xy}\nonumber\\
1&=&X_{\bp_{yz}}\times \left(\prod_{\bl_a\subset \bp_{yz}}\tau^Z_{\bl_a}\right)\times \left(\mu^X_{\bp_{yz}-\ex}\right)^{\hx+1}
=
\vcenter{\hbox{
    \begin{tikzpicture}[tdplot_main_coords]
        \draw[dashed](0,0,0)--++(0,0,1)--++(0,1,0)--++(0,0,-1)--cycle;
        \fill[red](0,.325,.325)--++(0,0,.25)--++(0,.25,0)--++(0,0,-.25)--cycle;
        \node at(0,.325,.625)[red]{$X$};
        \foreach\i in{0,1}{
            \fill(0,.5,\i)circle(.1);
            \fill(0,\i,.5)circle(.1);
        }
        \node at(0,.5,0)[left]{$\tau^Z$};
        \fill[blue](-.5,.5,.5)circle(.1)node[left]{$(\mu^X)^{\hat{x}+1}$};
    \end{tikzpicture}
}}
(=\vcentcolon G^X_{\bp_{yz}})\quad\forall\bp_{yz}\nonumber\\
1&=&X_{\bp_{zx}}\times \left(\prod_{\bl_a\subset \bp_{zx}}\tau^Z_{\bl_a}\right)\times \left(\sigma^X_{\bp_{zx}-\ex}\right)^{\hx}
=
\vcenter{\hbox{
    \begin{tikzpicture}[tdplot_main_coords]
        \draw[dashed](0,0,0)--++(0,0,1)--++(1,0,0)--++(0,0,-1)--cycle;
        \fill[red](.325,0,.325)--++(0,0,.25)--++(.25,0,0)--++(0,0,-.25)--cycle;
        \node at(.325,0,.625)[red]{$X$};
        \foreach\i in{0,1}{
            \fill(.5,0,\i)circle(.1);
            \fill(\i,0,.5)circle(.1);
        }
        \node at(1,.5,0)[below]{$\tau^Z$};
        \draw[green](0,0,.5)circle(.15)node[left]{$(\sigma^X)^{\hat{x}}$};
    \end{tikzpicture}
}}
(=\vcentcolon G^X_{\bp_{zx}})\quad\forall\bp_{zx}.\label{gauss_final}
\end{eqnarray}
Aside from the $x$-ordinate dependent terms, i.e., the ones which have the form ``$(\bullet)^{\hx}$'' or ``$(\bullet)^{\hx+1}$'',  the first [last] three conditions are imposed for gauging one [the other] $1$-form symmetry~\eqref{mem1}[\eqref{mem2}].  The $x$-ordinate dependent terms are introduced so that the all the Gauss's laws are commute. 
Moreover, product of the position dependent terms is subject to the following conditions:
\begin{eqnarray}
    \prod_{\hx=1}^{L_x}\prod_{\hy=1}^{L_y}\left(\tau^X_{(\hx+1,\hy,\hz+\frac{1}{2})}\right)^{\hx+1}=\prod_{\hx=1}^{L_x}\prod_{\hy=1}^{L_y}\left(\sigma^X_{(\hx,\hy+\frac{1}{2},\hz)}\right)^{\hx}=1\quad\forall\hz\nonumber\\
     \prod_{\hy=1}^{L_y}\prod_{\hz=1}^{L_z}\left(\tau^X_{(\hx+\frac{1}{2},\hy,\hz)}\right)^{\hx}=\prod_{\hy=1}^{L_y}\prod_{\hz=1}^{L_z}\left(\mu^X_{(\hx-\frac{1}{2},\hy+\frac{1}{2},\hz+\frac{1}{2})}\right)^{\hx+1}=1\quad\forall\hx\nonumber\\
     \prod_{\hz=1}^{L_z}\prod_{\hx=1}^{L_x}\left(\tau^X_{(\hx+1,\hy+\frac{1}{2},\hz)}\right)^{\hx+1}=\prod_{\hz=1}^{L_z}\prod_{\hx=1}^{L_x}\left(\sigma^X_{(\hx,\hy,\hz+\frac{1}{2})}\right)^{\hx}=1\quad\forall\hy.\label{condition}
\end{eqnarray}
These conditions~\eqref{condition} ensure that the global symmetries~\eqref{mem1}\eqref{mem2} are decomposed into local ones via,  $\prod_{\hx=1}^{L_x}\prod_{\hy=1}^{L_y}G^X_{\bp_{xy}}|_{\hz=1}=U^{X(1)}_{xy}$, and similarly for other symmetries, which is consistent with the fact that gauging is a procedure to promote a global to a local ones.
\par
Similar to the 2D case~(Sec.~\ref{sec3}), we minimally couple the terms to the symmetries, and add gauge invariant fluxes to arrive at the following gauged Hamiltonian:
\begin{eqnarray}
    H_{3D/\mathbb{Z}_2^{(1)}\times\mathbb{Z}^{(1)}_2}=
    &-&J_1\sum_{\bl_{a}}\left(\prod_{ \bp_{a^\prime b^\prime}\supset\bl_{a}}Z_{\bp_{a^\prime b^\prime}}\times \tau^X_{\bl_a}\right)-J_3\sum_{\bl_y}\left(\prod_{\bp_{xy}\supset \bl_y}X_{\bp_{xy}}\times\prod_{\br\subset \bl_y}X_\br\times\sigma^Z_{\bl_y}\right)  \nonumber\\
    &-&J_4\sum_{\bl_z}\left(\prod_{\bp_{zx}\supset \bl_z}X_{\bp_{zx}}\times\prod_{\br\subset \bl_z}X_\br\times\sigma^Z_{\bl_z}\right)
    -J_5\sum_\bc\left(\prod_{\bp_{xy}\subset \bc}X_{\bp_{xy}}\times\prod_{\bp_{zx}\subset \bc}X_{\bp_{zx}}\times\mu^Z_{\bc}\right)\nonumber\\
    & -&J_M\sum_\br\prod_{\bl_a\supset\br}\tau^X_{\bl_a}-J_M^\prime\sum_{\bp_{yz}}\left(\prod_{\bl_a\subset\bp_{yz}}\sigma^X_{\bl_a}\times\prod_{\bc\supset\bp_{yz}}\mu^X_{\bc}\right)\label{ghami3}
\end{eqnarray}
with $J_M,J^\prime_M\to\infty$. 
Graphically, it is described by
\begin{eqnarray}
    H_{3D/\mathbb{Z}_2^{(1)}\times\mathbb{Z}_2^{(1)}}
    &=&
    -J_1\qty(
        \sum_{\bx}
        \vcenter{\hbox{
            \begin{tikzpicture}[tdplot_main_coords]
                \draw[dashed](0,-1,0)--(0,1,0)--(1,1,0)--(1,-1,0)--cycle;
                \draw[dashed](0,0,-1)--(0,0,1)--(1,0,1)--(1,0,-1)--cycle;
                \draw[dashed](0,0,0)--(1,0,0);
                \fill(.5,0,0)circle(.1)node[above]{$\tau^X$};
                \fill[red, opacity=.5](.375,-.375,0)--(.625,-.375,0)--(.625,-.625,0)--(.375,-.625,0)--cycle;
                \fill[red, opacity=.5](.375,.375,0)--(.625,.375,0)--(.625,.625,0)--(.375,.625,0)--cycle;
                \fill[red, opacity=.5](.375,0,-.375)--(.625,0,-.375)--(.625,0,-.625)--(.375,0,-.625)--cycle;
                \fill[red, opacity=.5](.375,0,.375)--(.625,0,.375)--(.625,0,.625)--(.375,0,.625)--cycle;
                \node at(.5,-1,0)[red]{$Z$};
            \end{tikzpicture}
        }}
        +
        \sum_{\by}
        \vcenter{\hbox{
            \begin{tikzpicture}[tdplot_main_coords]
                \draw[dashed](-1,0,0)--(1,0,0)--(1,1,0)--(-1,1,0)--cycle;
                \draw[dashed](0,0,-1)--(0,0,1)--(0,1,1)--(0,1,-1)--cycle;
                \draw[dashed](0,0,0)--(0,1,0);
                \fill(0,.5,0)circle(.1)node[right]{$\tau^X$};
                \fill[red, opacity=.5](-.375,.375,0)--(-.625,.375,0)--(-.625,.625,0)--(-.375,.625,0)--cycle;
                \fill[red, opacity=.5](.375,.375,0)--(.625,.375,0)--(.625,.625,0)--(.375,.625,0)--cycle;
                \fill[red, opacity=.5](0,.375,-.375)--(0,.625,-.375)--(0,.625,-.625)--(0,.375,-.625)--cycle;
                \fill[red, opacity=.5](0,.375,.375)--(0,.625,.375)--(0,.625,.625)--(0,.375,.625)--cycle;
                \node at(.5,-1,0)[red]{$Z$};
            \end{tikzpicture}
        }}
        +
        \sum_{\bz}
        \vcenter{\hbox{
            \begin{tikzpicture}[tdplot_main_coords]
                \draw[dashed](-1,0,0)--(1,0,0)--(1,0,1)--(-1,0,1)--cycle;
                \draw[dashed](0,-1,0)--(0,1,0)--(0,1,1)--(0,-1,1)--cycle;
                \draw[dashed](0,0,0)--(0,0,1);
                \fill(0,0,.5)circle(.1)node[below right]{$\tau^X$};
                \fill[red, opacity=.5](-.375,0,.375)--(-.625,0,.375)--(-.625,0,.625)--(-.375,0,.625)--cycle;
                \fill[red, opacity=.5](.375,0,.375)--(.625,0,.375)--(.625,0,.625)--(.375,0,.625)--cycle;
                \fill[red, opacity=.5](0,-.375,.375)--(0,-.625,.375)--(0,-.625,.625)--(0,-.375,.625)--cycle;
                \fill[red, opacity=.5](0,.375,.375)--(0,.625,.375)--(0,.625,.625)--(0,.375,.625)--cycle;
                \node at(.5,0,1)[red]{$Z$};
            \end{tikzpicture}
        }}
    )
    \nonumber
    \\
    &&
    -J_3\sum_{\bm{l}_y}
    \vcenter{\hbox{
        \begin{tikzpicture}[tdplot_main_coords]
            \draw[dashed](-1,0,0)--(1,0,0)--(1,1,0)--(-1,1,0)--cycle;
            \draw[dashed](0,0,0)--(0,1,0);
            \fill[green](0,.5,0)circle(.1)node[left]{$\sigma^X$};
            \fill[red, opacity=.5](-.375,.375,0)--(-.625,.375,0)--(-.625,.625,0)--(-.375,.625,0)--cycle;
            \fill[red, opacity=.5](.375,.375,0)--(.625,.375,0)--(.625,.625,0)--(.375,.625,0)--cycle;
            \node at(1,.5,0)[red]{$X$};
            \fill[blue](0,0,0)circle(.1)node[below]{$X$};
            \fill[blue](0,1,0)circle(.1)node[above]{$X$};
        \end{tikzpicture}
    }}
    -J_4\sum_{\bm{l}_z}
    \vcenter{\hbox{
        \begin{tikzpicture}[tdplot_main_coords]
            \draw[dashed](-1,0,0)--(1,0,0)--(1,0,1)--(-1,0,1)--cycle;
            \draw[dashed](0,0,0)--(0,0,1);
            \fill[green](0,0,.5)circle(.1)node[right]{$\sigma^X$};
            \fill[red, opacity=.5](-.375,0,.375)--(-.625,0,.375)--(-.625,0,.625)--(-.375,0,.625)--cycle;
            \fill[red, opacity=.5](.375,0,.375)--(.625,0,.375)--(.625,0,.625)--(.375,0,.625)--cycle;
            \node at(1,0,.5)[red]{$X$};
            \fill[blue](0,0,0)circle(.1)node[left]{$X$};
            \fill[blue](0,0,1)circle(.1)node[right]{$X$};
        \end{tikzpicture}
    }}
    -J_5\sum_{\bm{c}}
    \vcenter{\hbox{
        \begin{tikzpicture}[tdplot_main_coords]
            \draw[dashed](0,0,0)--(1,0,0)--(1,1,0)--(0,1,0)--cycle;
            \draw[dashed](0,0,0)--(0,0,1)--(0,1,1)--(0,1,0)--cycle;
            \draw[dashed](0,0,0)--(1,0,0)--(1,0,1)--(0,0,1)--cycle;
            \draw[dashed](1,0,0)--(1,1,0)--(1,1,1)--(1,0,1)--cycle;
            \draw[dashed](0,1,0)--(1,1,0)--(1,1,1)--(0,1,1)--cycle;
            \draw[dashed](0,0,1)--(1,0,1)--(1,1,1)--(0,1,1)--cycle;
            \fill[red, opacity=.5](.375,.375,0)--(.625,.375,0)--(.625,.625,0)--(.375,.625,0)--cycle;
            \fill[red, opacity=.5](.375,0,.375)--(.625,0,.375)--(.625,0,.625)--(.375,0,.625)--cycle;
            \fill[red, opacity=.5](.375,.375,1)--(.625,.375,1)--(.625,.625,1)--(.375,.625,1)--cycle;
            \fill[red, opacity=.5](.375,1,.375)--(.625,1,.375)--(.625,1,.625)--(.375,1,.625)--cycle;
            \node at(.5,1,1)[red]{$X$};
            \fill[blue](.5,.5,.5)circle(.1)node[right]{$\mu^X$};
        \end{tikzpicture}
    }}
    \nonumber
    \\
    &&
    -J_M\sum_{\bm{r}}
    \vcenter{\hbox{
        \begin{tikzpicture}[tdplot_main_coords]
            \draw[dashed](-1,0,0)--(1,0,0);
            \draw[dashed](0,-1,0)--(0,1,0);
            \draw[dashed](0,0,-1)--(0,0,1);
            \foreach\i in{1,-1}{
                \fill(\i/2,0,0)circle(.1);
                \fill(0,\i/2,0)circle(.1);
                \fill(0,0,\i/2)circle(.1);
            }
            \node at(0,0,-.5)[below left]{$\tau^X$};
        \end{tikzpicture}
    }}
    -J_M'\sum_{\bm{p}_{yz}}
    \vcenter{\hbox{
        \begin{tikzpicture}[tdplot_main_coords]
            \draw[dashed](0,0,0)--++(0,0,1)--++(0,1,0)--++(0,0,-1)--cycle;
            \foreach\i in{0,1}{
                \fill[green](0,.5,\i)circle(.1);
                \fill[green](0,\i,.5)circle(.1);
            }
            \node at(0,.5,0)[left,green]{$\sigma^X$};
            \fill[blue](-.5,.5,.5)circle(.1)node[left]{$\mu^X$};
            \fill[blue](.5,.5,.5)circle(.1)node[right]{$\mu^X$};
        \end{tikzpicture}
    }}
\end{eqnarray}
The gauged Hamiltonian~\eqref{ghami3} respects dual $1$-form symmetries, described by the following membrane operators:
\begin{eqnarray}
\mathcal{U}^{X(1)}_{xy}\vcentcolon=\prod_{\hx=1}^{L_x}\prod_{\hy=1}^{L_y}\tau^X_{(\hx,\hy,\frac{3}{2})},\quad \mathcal{U}^{X(1)}_{yz}\vcentcolon=\prod_{\hy=1}^{L_y}\prod_{\hz=1}^{L_z}\tau^X_{(\frac{3}{2}\hy,\hz)},\quad \mathcal{U}^{X(1)}_{zx}\vcentcolon=\prod_{\hz=1}^{L_z}\prod_{\hx=1}^{L_x}\tau^X_{(\hx,\frac{3}{2},\hz)}\nonumber\\
\mathcal{V}^{Z(1)}_{xy}\vcentcolon=\prod_{\hx=1}^{L_x}\prod_{\hy=1}^{L_y}\sigma^X_{(\hx,\hy+\frac{1}{2},1)},\quad \mathcal{V}^{Z(1)}_{yz}\vcentcolon=\prod_{\hy=1}^{L_y}\prod_{\hz=1}^{L_z}\mu^X_{(\frac{3}{2},\hy+\frac{1}{2},\hz+\frac{1}{2})},\quad \mathcal{V}^{X(1)}_{zx}\vcentcolon=\prod_{\hz=1}^{L_z}\prod_{\hx=1}^{L_x}\sigma^X_{(\hx,1,\hz+\frac{1}{2})}.\label{dual1}
\end{eqnarray}
After gauging, 
the translation operator in the $x$-direction is not gauge invariant, which can be seen from~e.g., $T_x G^{Z(1)}_{\bp_{xy}}T_x^{-1}\neq G^{Z(1)}_{\bp_{xy}}$ 
whereas the ones in $y$- and~$z$-direction are gauge invariant. To remedy this issue, we modify the lattice translation operator in the $x$-direction so that it is dressed with the TST transformation. To wit, 
\begin{eqnarray}
    T_x\to T_xW~(\vcentcolon=\mathcal{T}_x),\label{114}
\end{eqnarray}
where,
\begin{eqnarray}
    W=\mathcal{U}_{CZ}\times\mathcal{U}_D\times\mathcal{U}_{CZ}.\label{ucdcz}
\end{eqnarray}
Here, 
\begin{eqnarray}
    \mathcal{U}_{CZ}=\prod_{s=1}^{6}\mathcal{U}_{CZ,s}
\end{eqnarray}
with
\begin{eqnarray}
    \mathcal{U}_{CZ,1}\vcentcolon&=&\prod_{\bl_x}\exp\left[i\frac{\pi}{4}(1-\tau^Z_{\bl_x})\times\sum_{_{\bl_a\supset\bl_x-\ex}}(1-\sigma^Z_{\bl_a})\right],\nonumber\\
     \mathcal{U}_{CZ,2}\vcentcolon&=&\prod_{\bl_y}\exp\left[\frac{i\pi}{4}(1-\tau^Z_{\bl_y})\times\left\{\sum_{s=0,1}(1-\sigma^Z_{\bl_y-s\exx})+\sum_{t=\pm1}(1-\mu^Z_{\bl_y-\ex+t\ez})\right\}\right],\nonumber\\
      \mathcal{U}_{CZ,3}\vcentcolon&=&\prod_{\bl_z}\exp\left[\frac{i\pi}{4}(1-\tau^Z_{\bl_z})\times\left\{\sum_{s=0,1}(1-\sigma^Z_{\bl_z-s\exx})+\sum_{t=\pm1}(1-\mu^Z_{\bl_z-\ex+t\ey})\right\}\right],\nonumber\\
\mathcal{U}_{CZ,4}\vcentcolon&=&\prod_{\bl_y}\exp\left[\frac{i\pi}{4}(1-\sigma^Z_{\bl_y})\times\left\{\sum_{s=0,1}(1-\tau^Z_{\bl_y+s\exx})+\sum_{t=\pm1}(1-\tau^Z_{\bl_y+\ex+t\ey})\right\}\right],\nonumber\\
\mathcal{U}_{CZ,5}\vcentcolon&=&\prod_{\bl_z}\exp\left[\frac{i\pi}{4}(1-\sigma^Z_{\bl_z})\times\left\{\sum_{s=0,1}(1-\tau^Z_{\bl_z+s\exx})+\sum_{t=\pm1}(1-\tau^Z_{\bl_z+\ex+t\ez})\right\}\right],\nonumber\\
\mathcal{U}_{CZ,6}\vcentcolon&=&\prod_{\bc}\exp\left[\frac{i\pi}{4}(1-\mu^Z_{\bc})\times\left\{\sum_{s=\pm1}(1-\tau^Z_{\bc+\exx+s\ey})+\sum_{t=\pm1}(1-\tau^Z_{\bc+\ex+t\ez})\right\}\right].
\end{eqnarray}
These are product of control phase gate operators. Physically,  $\mathcal{U}_{CZ}$ implements stacking of a SPT phase protected by $\mathbb{Z}_2^{(1)}\times\mathbb{Z}_2^{(1)}$ symmetry~\cite{raussendorf2005long}.
Also,
\begin{eqnarray}
    \mathcal{U}_{D}\vcentcolon=\frac{1}{2} \sum_{\substack{\{\tau_{\bl_a}\},\{\sigma_{\bl_a}\},\{\mu_{\bc}\} \\ \{\tau^\prime_{\bl_a}\},\{\sigma^\prime_{\bl_a}\},\{\mu^\prime_{\bc}\} }}\Omega\times\Omega^\prime\ket{\{\tau^\prime_{\bl_a}\},\{\sigma^\prime_{\bl_a}\},\{\mu^\prime_\bc\}}\bra{\{\tau_{\bl_a}\},\{\sigma_{\bl_a}\},\{\mu_\bc\}}.\label{ud}
\end{eqnarray}
Here, $\ket{\{\tau^\prime_{\bl_a}\},\{\sigma^\prime_{\bl_a}\},\{\mu^\prime_\bc\}}$~$(\tau_{\bl_a},\sigma_{\bl_{a}},\mu_{\bc}=0,1)$ represents product of diagonal basis for Pauli operators, $\tau^Z_{\bl_a}$, $\sigma^Z_{\bl_a}$, $\mu_{\bc}$ on entire links and centers of cubes, satisfying
\begin{eqnarray}
   \tau^Z_{\bl_{a0}}\ket{\{\tau_{\bl_a}\},\{\sigma_{\bl_a}\},\{\bc_\bp\}}&=&(-1)^{\tau_{\bl_{a0}}}\ket{\{\tau_{\bl_a}\},\{\sigma_{\bl_a}\},\{\bc_\bp\}},\nonumber\\
    \tau^X_{\bl_{a0}}\ket{\{\tau_{\bl_a}\},\{\sigma_{\bl_a}\},\{\bc_\bp\}}&=&\ket{\{\tau_{\bl_{a0}\neq\bl_a};\tau_{\bl_{a0}}+1\},\{\sigma_{\bl_a}\},\{\bc_\bp\}},
\end{eqnarray}
and similarly for actions of other operators on the state. 
A phase factor $\Omega$ in~\eqref{ud} is described by
\begin{eqnarray}
    \Omega\vcentcolon=\exp\left[i\pi\sum_{\bl_x}\left(\sum_{\bl_a\supset \bl_x-\ex}\sigma^\prime_{\bl_a}\right)\tau_{\bl_x}+i\pi\sum_{\bl_y}\sum_{s=0,1}\sum_{t=\pm1}\left(\sigma^\prime_{\bl_y-s\exx}+\mu^\prime_{\bl_y-\ex+t\ez}\right)\tau_{\bl_y}\right]\nonumber\\
    \times\exp\left[i\pi\sum_{\bl_z}\sum_{s=0,1}\sum_{t=\pm1}\left(\sigma^\prime_{\bl_z-s\exx}+\mu^\prime_{\bl_x-\ex+t\ey}\right)\tau_{\bl_z}\right],\label{om}
\end{eqnarray}
whereas $\Omega^\prime$ is obtained from~\eqref{om} by switching variables with prime and those without prime. 
The operator $\mathcal{U}_D$ is the duality mapping between local operators via gauging $\mathbb{Z}_2^{(1)} \times \mathbb{Z}_2^{(1)}$ \footnote{A tensor network representation of a duality mapping via gauging $\mathbb{Z}_2^{(1)}$ is also known in Ref.~\cite{Gorantla:2024ocs}. Note that our operator $\mathcal{U}_D$ corresponds to a duality mapping via gauging \emph{two} $1$-form $\mathbb{Z}_2$ symmetries.}.  \par 
The operator $W$ in~\eqref{ucdcz} is the 3D analog of the TST transformation that maps between an SPT phase and SSB state protected by 
$\mathbb{Z}_2^{(1)}\times\mathbb{Z}_2^{(1)}$ symmetry~\eqref{dual1}. 
For instance, one finds that  
\begin{eqnarray}
    W  \left(\prod_{\bl_a\subset \bp_{yz}}\tau^Z_{\bl_a}\right) =  \left(\prod_{\bl_a\subset \bp_{yz}}\tau^Z_{\bl_a}\times\sigma^X_{\bp_{xy}-\ex} \right)W.
\end{eqnarray}
Furthermore, by resorting to the similar argument presented in Appendix.~\ref{app22}, after some algebra, we have 
\begin{eqnarray}
   \mathcal{U}_{D}\times \mathcal{U}_{D}=\left[\frac{1}{2}\sum_{ab=xy,yz,zx}(1+\mathcal{U}^{X(1)}_{ab})\right]\times \left[\frac{1}{2}\sum_{ab=xy,yz,zx}(1+\mathcal{V}^{Z(1)}_{ab})\right],\label{conde}
\end{eqnarray}
implying that fusion between the translation operators yields condensation defects of the dual~$1$-form symmetries~\eqref{dual1}.
\par
The modified translation operator is gauge invariant, which is verified by $\mathcal{T}_x G^{Z}_{\bp_{xy}}\mathcal{T}_x^{-1}=G^{Z}_{\bp_{xy}+\exx}$, and similarly for other Gauss's laws~\eqref{gauss_final}. 
Further, from~\eqref{conde}, jointly with the fact that $\mathcal{U}_{CZ}^2=1$, 
it follows that 
\begin{eqnarray}
    \mathcal{T}_x\times \mathcal{T}_x^{-1}=\left[\frac{1}{2}\sum_{ab=xy,yz,zx}(1+\mathcal{U}^{X(1)}_{ab})\right]\times \left[\frac{1}{2}\sum_{ab=xy,yz,zx}(1+\mathcal{V}^{Z(1)}_{ab})\right]~\mathbb{I}.
\end{eqnarray}
Hence, the translation operator becomes non-invertible. 
\section{Derivation of fusion rules} \label{appendix.E}
In this appendix, we provide the derivation of \eqref{fusion in field theory}. 
\begin{itemize}
    \item $\mathcal{N} \times \eta_a = \eta_a \times \mathcal{N} = \mathcal{N}$
\end{itemize}
\allowdisplaybreaks
    \begin{align*}
        &\mathcal{N} \times \eta_a \\
        =& \sum_{\substack{\phi_a^{(p)}\in C^{p}(Y,\mathbb{Z}_2) \\ \phi_b^{(d-p-1)}\in C^{d-p-1}(Y,\mathbb{Z}_2)}} g(Y) \exp\Bigg[\frac{2\pi i}{2} \int_Y\phi_b^{(d-p-1)}a^{(p+1)}+\phi_a^{(p)}b^{(d-p)}+\phi_a^{(p)}\delta\phi_b^{(d-p-1)}\Bigg] \times  \exp\Big(i\pi\int_{\Sigma_{p+1}} a^{(p+1)}\Big) \\
        =&\sum_{\substack{\phi_a^{(p)}\in C^{p}(Y,\mathbb{Z}_2) \\ \phi_b^{(d-p-1)}\in C^{d-p-1}(Y,\mathbb{Z}_2)}} g(Y) \exp\Bigg[\frac{2\pi i}{2} \int_Y\big(\phi_b^{(d-p-1)}+\mathrm{PD}(\Sigma_{p+1})\big)a^{(p+1)}+\phi_a^{(p)}b^{(d-p)}+\phi_a^{(p)}\delta\phi_b^{(d-p-1)}\Bigg] \\
        =&\sum_{\substack{\phi_a^{(p)}\in C^{p}(Y,\mathbb{Z}_2) \\ \phi_b^{(d-p-1)}\in C^{d-p-1}(Y,\mathbb{Z}_2)}} g(Y) \exp\Bigg[\frac{2\pi i}{2} \int_Y\phi_b^{(d-p-1)}a^{(p+1)}+\phi_a^{(p)}b^{(d-p)}+\phi_a^{(p)}\delta\phi_b^{(d-p-1)}\Bigg] \\
        =&~\mathcal{N}
    \end{align*}
In the second equality, we rewrite the Wilson surface $\eta_a$ using the Poincar\'e duality, that is,
\begin{equation}
    \exp\Big(i\pi\int_{\Sigma_{p+1}} a^{(p+1)}\Big) = \exp\Big(i\pi\int_{M} \mathrm{PD}(\Sigma_{p+1})\cup a^{(p+1)}\Big),
\end{equation}
where $\Sigma_{p+1}$ is $(p+1)$-dimensional closed surface and $\mathrm{PD}(\Sigma_{p+1})$ denotes Poincaré dual of $\Sigma_{p+1}$ on $Y$. In the third equality, we redefine $\phi_b^{(d-p-1)}$ by performing the shift $\phi_b^{(d-p-1)} \rightarrow \phi_b^{(d-p-1)} + \mathrm{PD}(\Sigma_{p+1})$. A similar argument applies to $\eta_a \times \mathcal{N}$.

\begin{itemize}
    \item $\mathcal{N} \times \eta_b = \eta_b \times \mathcal{N} = \mathcal{N}$
\end{itemize}
This fusion rule can be shown by a similar calculation as above.

\begin{itemize}
    \item $\mathcal{N} \times \mathcal{N}^{\dagger} \sim C_aC_b$
\end{itemize}
\allowdisplaybreaks
    \begin{align*}
        &\mathcal{N} \times \mathcal{N}^{\dagger} \\
       \sim& \sum_{\substack{\phi_a^{(p)}, \phi_a'^{(p)}\in C^{p}(Y,\mathbb{Z}_2) \\ \phi_b^{(d-p-1)}, \phi_b'^{(d-p-1)}\in C^{d-p-1}(Y,\mathbb{Z}_2)}} \exp\Bigg[\frac{2\pi i}{2} \int_Y\big(\phi_b^{(d-p-1)} + \phi_b'^{(d-p-1)}\big)a^{(p+1)}+\big(\phi_a^{(p)} + \phi_a'^{(p)} \big) b^{(d-p)} \Bigg] \\
       & \hspace{4cm}\times \exp\Bigg[\frac{2\pi i}{2} \int_Y \phi_a^{(p)}\delta\phi_b^{(d-p-1)} + \phi_a'^{(p)}\delta\phi_b'^{(d-p-1)} \Bigg] \\
       \sim & \sum_{\substack{\phi_a^{(p)}, \bar{\phi}_a^{(p)}\in C^{p}(Y,\mathbb{Z}_2) \\ \phi_b^{(d-p-1)}, \bar{\phi}_b^{(d-p-1)}\in C^{d-p-1}(Y,\mathbb{Z}_2)}} \exp\Bigg[\frac{2\pi i}{2} \int_Y\bar{\phi}_b^{(d-p-1)}a^{(p+1)}+\bar{\phi}_a^{(p)} b^{(d-p)} \Bigg] \\
       & \hspace{4cm}\times \exp\Bigg[\frac{2\pi i}{2} \int_Y \bar{\phi}_a^{(p)}\delta\bar{\phi}_b^{(d-p-1)} + \bar{\phi}_a^{(p)}\delta\phi_b^{(d-p-1)} + \phi_a^{(p)}\delta\bar{\phi}_b^{(d-p-1)} \Bigg] \\
       \sim&  \sum_{\substack{\bar{\phi}_a^{(p)}\in H^{p}(Y,\mathbb{Z}_2) \\ \bar{\phi}_b^{(d-p-1)}\in H^{d-p-1}(Y,\mathbb{Z}_2)}} \exp\Bigg[\frac{2\pi i}{2} \int_Y\bar{\phi}_b^{(d-p-1)}a^{(p+1)}+\bar{\phi}_a^{(p)} b^{(d-p)} \Bigg] \\
     \sim & \sum_{\substack{\Sigma_{p+1}\in H_{p+1}(Y,\mathbb{Z}_2) \\ \Sigma_{(d-p)}\in H_{d-p}(Y,\mathbb{Z}_2)}}\eta_a(\Sigma_{p+1}) \times \eta_b(\Sigma_{d-p}) \\
     \sim &~C_aC_b
    \end{align*}
In the second line, we introduce $\bar{\phi}_a:=\phi_a +\phi_a'$ and $\bar{\phi}_b:=\phi_b +\phi_b'$. In the third line, we performe the integral over $\phi_a$ and $\phi_b$, which imposes that $\bar{\phi}_a$ and $\bar{\phi}_b$ are $\mathbb{Z}_2$-valued cocycles. In the next line, using the Poincaré duality, we rewrite it as
\begin{eqnarray}
    \exp\Big(\frac{2\pi i}{2}\int_Y\bar{\phi}_b^{(d-p-1)}a^{(p+1)}\Big) = \exp\Big(\frac{2\pi i}{2}\int_{\mathrm{PD}(\bar{\phi}_b^{(d-p-1)})}a^{(p+1)}\Big) =
    \eta_a(\Sigma_{p+1}), 
\end{eqnarray}
 where $\Sigma_{p+1}:=\mathrm{PD}(\bar{\phi}_b^{(d-p-1)})$. The same calculation applies to $\eta_b(\Sigma_{d-p})$.
\end{appendix}

\bibliography{main}

@article{oishi2026type,
  title={Type-IV't Hooft Anomalies on the Lattice: Emergent Higher-Categorical Symmetries and Applications to LSM Systems},
  author={Oishi, Tsubasa and Ebisu, Hiromi},
  journal={arXiv preprint arXiv:2604.02856},
  year={2026}
}

@article{Ebisu:2024eew,
    author = "Ebisu, Hiromi and Honda, Masazumi and Nakanishi, Taiichi",
    title = "{Anomaly inflow for dipole symmetry and higher form foliated field theories}",
    eprint = "2406.04919",
    archivePrefix = "arXiv",
    primaryClass = "cond-mat.str-el",
    reportNumber = "RIKEN-iTHEMS-Report-24",
    doi = "10.1007/JHEP09(2024)061",
    journal = "JHEP",
    volume = "09",
    pages = "061",
    year = "2024"
}

@article{wang2015field,
  title={Field-theory representation of gauge-gravity symmetry-protected topological invariants, group cohomology, and beyond},
  author={Wang, Juven C and Gu, Zheng-Cheng and Wen, Xiao-Gang},
  journal={Physical review letters},
  volume={114},
  number={3},
  pages={031601},
  year={2015},
  publisher={APS}
}

@article{raussendorf2005long,
  title={Long-range quantum entanglement in noisy cluster states},
  author={Raussendorf, Robert and Bravyi, Sergey and Harrington, Jim},
  journal={Physical Review A—Atomic, Molecular, and Optical Physics},
  volume={71},
  number={6},
  pages={062313},
  year={2005},
  publisher={APS}
}

@article{Gorantla:2022pii,
    author = "Gorantla, Pranay and Lam, Ho Tat and Seiberg, Nathan and Shao, Shu-Heng",
    title = "{Gapped lineon and fracton models on graphs}",
    eprint = "2210.03727",
    archivePrefix = "arXiv",
    primaryClass = "cond-mat.str-el",
    reportNumber = "YITP-SB-2022-33, MIT/CTP-5471",
    doi = "10.1103/PhysRevB.107.125121",
    journal = "Phys. Rev. B",
    volume = "107",
    number = "12",
    pages = "125121",
    year = "2023"
}

@article{Seiberg:2024wgj,
    author = "Seiberg, Nathan",
    title = "{Ferromagnets, a new anomaly, instantons, and (noninvertible) continuous translations}",
    eprint = "2406.06698",
    archivePrefix = "arXiv",
    primaryClass = "cond-mat.str-el",
    doi = "10.21468/SciPostPhys.18.2.063",
    journal = "SciPost Phys.",
    volume = "18",
    number = "2",
    pages = "063",
    year = "2025"
}

@article{Seiberg:2024yig,
    author = "Seiberg, Nathan",
    title = "{Anomalous continuous translations}",
    eprint = "2412.14434",
    archivePrefix = "arXiv",
    primaryClass = "cond-mat.str-el",
    doi = "10.21468/SciPostPhys.19.2.031",
    journal = "SciPost Phys.",
    volume = "19",
    number = "2",
    pages = "031",
    year = "2025"
}

@phdthesis{deWildPropitius:1995cf,
    author = "de Wild Propitius, Mark Dirk Frederik",
    title = "{Topological interactions in broken gauge theories}",
    eprint = "hep-th/9511195",
    archivePrefix = "arXiv",
    school = "Amsterdam U.",
    year = "1995"
}

@article{bulmash2025defect,
  title={Defect networks for topological phases protected by modulated symmetries},
  author={Bulmash, Daniel},
  journal={arXiv preprint arXiv:2508.06604},
  year={2025}
}

@article{PhysRevX.10.011047,
  title = {Ergodicity Breaking Arising from Hilbert Space Fragmentation in Dipole-Conserving Hamiltonians},
  author = {Sala, Pablo and Rakovszky, Tibor and Verresen, Ruben and Knap, Michael and Pollmann, Frank},
  journal = {Phys. Rev. X},
  volume = {10},
  issue = {1},
  pages = {011047},
  numpages = {19},
  year = {2020},
  month = {Feb},
  publisher = {American Physical Society},
  doi = {10.1103/PhysRevX.10.011047},
  url = {https://link.aps.org/doi/10.1103/PhysRevX.10.011047}
}

@article{Gaiotto:2014kfa,
    author = "Gaiotto, Davide and Kapustin, Anton and Seiberg, Nathan and Willett, Brian",
    title = "{Generalized Global Symmetries}",
    eprint = "1412.5148",
    archivePrefix = "arXiv",
    primaryClass = "hep-th",
    doi = "10.1007/JHEP02(2015)172",
    journal = "JHEP",
    volume = "02",
    pages = "172",
    year = "2015"
}

@Article{10.21468/SciPostPhys.14.5.106,
	title={{Anisotropic higher rank $\mathbb{Z}_N$ topological phases on graphs}},
	author={Hiromi Ebisu and Bo Han},
	journal={SciPost Phys.},
	volume={14},
	pages={106},
	year={2023},
	publisher={SciPost},
	doi={10.21468/SciPostPhys.14.5.106},
	url={https://scipost.org/10.21468/SciPostPhys.14.5.106},
}

@article{Cobanera:2011wn,
    author = "Cobanera, Emilio and Ortiz, Gerardo and Nussinov, Zohar",
    title = "{The Bond-Algebraic Approach to Dualities}",
    eprint = "1103.2776",
    archivePrefix = "arXiv",
    primaryClass = "cond-mat.stat-mech",
    doi = "10.1080/00018732.2011.619814",
    journal = "Adv. Phys.",
    volume = "60",
    pages = "679--798",
    year = "2011"
}

@article{chamon,
  title = {Quantum Glassiness in Strongly Correlated Clean Systems: An Example of Topological Overprotection},
  author = {Chamon, Claudio},
  journal = {Phys. Rev. Lett.},
  volume = {94},
  issue = {4},
  pages = {040402},
  numpages = {4},
  year = {2005},
  month = {Jan},
  publisher = {American Physical Society},
  doi = {10.1103/PhysRevLett.94.040402},
  url = {https://link.aps.org/doi/10.1103/PhysRevLett.94.040402}
}

@article{yan2024generalized,
  title={Generalized Kramers-Wanier duality from bilinear phase map},
  author={Yan, Han and Li, Linhao},
  journal={arXiv preprint arXiv:2403.16017},
  year={2024}
}

@article{Haah2011,
  title = {Local stabilizer codes in three dimensions without string logical operators},
  author = {Haah, Jeongwan},
  journal = {Phys. Rev. A},
  volume = {83},
  issue = {4},
  pages = {042330},
  numpages = {16},
  year = {2011},
  month = {Apr},
  publisher = {American Physical Society},
  doi = {10.1103/PhysRevA.83.042330},
  url = {https://link.aps.org/doi/10.1103/PhysRevA.83.042330}
}

@article{Gorantla:2022eem,
    author = "Gorantla, Pranay and Lam, Ho Tat and Seiberg, Nathan and Shao, Shu-Heng",
    title = "{Global dipole symmetry, compact Lifshitz theory, tensor gauge theory, and fractons}",
    eprint = "2201.10589",
    archivePrefix = "arXiv",
    primaryClass = "cond-mat.str-el",
    doi = "10.1103/PhysRevB.106.045112",
    journal = "Phys. Rev. B",
    volume = "106",
    number = "4",
    pages = "045112",
    year = "2022"
}

@article{Ebisu_BO_2025,
  title = {Noninvertible operators in one, two, and three dimensions via gauging spatially modulated symmetry},
  author = {Ebisu, Hiromi and Han, Bo},
  journal = {Phys. Rev. B},
  volume = {111},
  issue = {3},
  pages = {035149},
  numpages = {24},
  year = {2025},
  month = {Jan},
  publisher = {American Physical Society},
  doi = {10.1103/PhysRevB.111.035149},
  url = {https://link.aps.org/doi/10.1103/PhysRevB.111.035149}
}

@article{Kapustin2014,
  title = {Anomalous Discrete Symmetries in Three Dimensions and Group Cohomology},
  author = {Kapustin, Anton and Thorngren, Ryan},
  journal = {Phys. Rev. Lett.},
  volume = {112},
  issue = {23},
  pages = {231602},
  numpages = {4},
  year = {2014},
  month = {Jun},
  publisher = {American Physical Society},
  doi = {10.1103/PhysRevLett.112.231602},
  url = {https://link.aps.org/doi/10.1103/PhysRevLett.112.231602}
}

@article{spt2013,
  title = {Symmetry protected topological orders and the group cohomology of their symmetry group},
  author = {Chen, Xie and Gu, Zheng-Cheng and Liu, Zheng-Xin and Wen, Xiao-Gang},
  journal = {Phys. Rev. B},
  volume = {87},
  issue = {15},
  pages = {155114},
  numpages = {48},
  year = {2013},
  month = {Apr},
  publisher = {American Physical Society},
  doi = {10.1103/PhysRevB.87.155114},
  url = {https://link.aps.org/doi/10.1103/PhysRevB.87.155114}
}

@article{Vijay,
  title = {Fracton topological order, generalized lattice gauge theory, and duality},
  author = {Vijay, Sagar and Haah, Jeongwan and Fu, Liang},
  journal = {Phys. Rev. B},
  volume = {94},
  issue = {23},
  pages = {235157},
  numpages = {9},
  year = {2016},
  month = {Dec},
  publisher = {American Physical Society},
  doi = {10.1103/PhysRevB.94.235157},
  url = {https://link.aps.org/doi/10.1103/PhysRevB.94.235157}
}

@article{KITAEV20032,
title = {Fault-tolerant quantum computation by anyons},
journal = {Annals of Physics},
volume = {303},
number = {1},
pages = {2-30},
year = {2003},
issn = {0003-4916},
doi = {https://doi.org/10.1016/S0003-4916(02)00018-0},
url = {https://www.sciencedirect.com/science/article/pii/S0003491602000180},
author = {A.Yu. Kitaev}
}

@article{griffin2015scalar,
    author = "Griffin, Tom and Grosvenor, Kevin T. and Horava, Petr and Yan, Ziqi",
    title = "{Scalar Field Theories with Polynomial Shift Symmetries}",
    eprint = "1412.1046",
    archivePrefix = "arXiv",
    primaryClass = "hep-th",
    doi = "10.1007/s00220-015-2461-2",
    journal = "Commun. Math. Phys.",
    volume = "340",
    number = "3",
    pages = "985--1048",
    year = "2015"
}

@ARTICLE{1986LMaPh..12...57A,
       author = {{Affleck}, Ian and {Lieb}, Elliott H.},
        title = "{A proof of part of Haldane's conjecture on spin chains}",
      journal = {Letters in Mathematical Physics},
         year = 1986,
        month = jul,
       volume = {12},
       number = {1},
        pages = {57-69},
          doi = {10.1007/BF00400304},
       adsurl = {https://ui.adsabs.harvard.edu/abs/1986LMaPh..12...57A}
}

@article{Antinucci:2025fjp,
    author = "Antinucci, Andrea and Copetti, Christian and Gai, Yuhan and Schafer-Nameki, Sakura",
    title = "{Categorical Anomaly Matching}",
    eprint = "2508.00982",
    archivePrefix = "arXiv",
    primaryClass = "hep-th",
    month = "8",
    year = "2025"
}

@article{Ebisu:2025mtb,
    author = "Ebisu, Hiromi and Han, Bo and Cao, Weiguang",
    title = "{Modulated symmetries from generalized Lieb-Schultz-Mattis anomalies}",
    eprint = "2510.18689",
    archivePrefix = "arXiv",
    primaryClass = "cond-mat.str-el",
    reportNumber = "RIKEN-iTHEMS-Report-25",
    month = "10",
    year = "2025"
}

@article{2024multipole,
  title = {Multipole and fracton topological order via gauging foliated symmetry protected topological phases},
  author = {Ebisu, Hiromi and Honda, Masazumi and Nakanishi, Taiichi},
  journal = {Phys. Rev. Res.},
  volume = {6},
  issue = {2},
  pages = {023166},
  numpages = {18},
  year = {2024},
  month = {May},
  publisher = {American Physical Society},
  doi = {10.1103/PhysRevResearch.6.023166},
  url = {https://link.aps.org/doi/10.1103/PhysRevResearch.6.023166}
}

@article{Pace:2025hpb,
    author = {Pace, Salvatore D. and Aksoy, {\"O}mer M. and Lam, Ho Tat},
    title = "{Spacetime symmetry-enriched SymTFT: from LSM anomalies to modulated symmetries and beyond}",
    eprint = "2507.02036",
    archivePrefix = "arXiv",
    primaryClass = "cond-mat.str-el",
    reportNumber = "MIT-CTP/5884",
    month = "7",
    year = "2025"
}

@article{Ebisu:2023idd,
    author = "Ebisu, Hiromi and Honda, Masazumi and Nakanishi, Taiichi",
    title = "{Foliated field theories and multipole symmetries}",
    eprint = "2310.06701",
    archivePrefix = "arXiv",
    primaryClass = "cond-mat.str-el",
    reportNumber = "YITP-23-129, RIKEN-iTHEMS-Report-23",
    doi = "10.1103/PhysRevB.109.165112",
    journal = "Phys. Rev. B",
    volume = "109",
    number = "16",
    pages = "165112",
    year = "2024"
}

@article{Cheng:2015kce,
    author = "Cheng, Meng and Zaletel, Michael and Barkeshli, Maissam and Vishwanath, Ashvin and Bonderson, Parsa",
    title = "{Translational Symmetry and Microscopic Constraints on Symmetry-Enriched Topological Phases: A View from the Surface}",
    eprint = "1511.02263",
    archivePrefix = "arXiv",
    primaryClass = "cond-mat.str-el",
    doi = "10.1103/PhysRevX.6.041068",
    journal = "Phys. Rev. X",
    volume = "6",
    number = "4",
    pages = "041068",
    year = "2016"
}

@article{Pretko:2018jbi,
    author = "Pretko, Michael",
    title = "{The Fracton Gauge Principle}",
    eprint = "1807.11479",
    archivePrefix = "arXiv",
    primaryClass = "cond-mat.str-el",
    doi = "10.1103/PhysRevB.98.115134",
    journal = "Phys. Rev. B",
    volume = "98",
    number = "11",
    pages = "115134",
    year = "2018"
}

@article{PhysRevLett.84.1535,
  title = {Commensurability, Excitation Gap, and Topology in Quantum Many-Particle Systems on a Periodic Lattice},
  author = {Oshikawa, Masaki},
  journal = {Phys. Rev. Lett.},
  volume = {84},
  issue = {7},
  pages = {1535--1538},
  numpages = {0},
  year = {2000},
  month = {Feb},
  publisher = {American Physical Society},
  doi = {10.1103/PhysRevLett.84.1535},
  url = {https://link.aps.org/doi/10.1103/PhysRevLett.84.1535}
}

@article{PhysRevB.69.104431,
  title = {Lieb-Schultz-Mattis in higher dimensions},
  author = {Hastings, M. B.},
  journal = {Phys. Rev. B},
  volume = {69},
  issue = {10},
  pages = {104431},
  numpages = {13},
  year = {2004},
  month = {Mar},
  publisher = {American Physical Society},
  doi = {10.1103/PhysRevB.69.104431},
  url = {https://link.aps.org/doi/10.1103/PhysRevB.69.104431}
}

@article{Lieb1961,
author = {Lieb, Elliott and Schultz, Theodore and Mattis, Daniel},
doi = {10.1016/0003-4916(61)90115-4},
journal = {Annals of Physics},
month = {dec},
number = {3},
pages = {407--466},
publisher = {Academic Press},
title = {{Two soluble models of an antiferromagnetic chain}},
url = {https://www.sciencedirect.com/science/article/pii/0003491661901154?via{\%}3Dihub},
volume = {16},
year = {1961}
}

@article{PhysRevB.86.115109,
  title = {Braiding statistics approach to symmetry-protected topological phases},
  author = {Levin, Michael and Gu, Zheng-Cheng},
  journal = {Phys. Rev. B},
  volume = {86},
  issue = {11},
  pages = {115109},
  numpages = {15},
  year = {2012},
  month = {Sep},
  publisher = {American Physical Society},
  doi = {10.1103/PhysRevB.86.115109},
  url = {https://link.aps.org/doi/10.1103/PhysRevB.86.115109}
}

@article{Seifnashri:2023dpa,
    author = "Seifnashri, Sahand",
    title = "{Lieb-Schultz-Mattis anomalies as obstructions to gauging (non-on-site) symmetries}",
    eprint = "2308.05151",
    archivePrefix = "arXiv",
    primaryClass = "cond-mat.str-el",
    doi = "10.21468/SciPostPhys.16.4.098",
    journal = "SciPost Phys.",
    volume = "16",
    number = "4",
    pages = "098",
    year = "2024"
}

@article{Pace:2024acq,
    author = {Pace, Salvatore D. and Lam, Ho Tat and Aksoy, {\"O}mer Mert},
    title = "{(SPT-)LSM theorems from projective non-invertible symmetries}",
    eprint = "2409.18113",
    archivePrefix = "arXiv",
    primaryClass = "cond-mat.str-el",
    reportNumber = "MIT-CTP/5772",
    doi = "10.21468/SciPostPhys.18.1.028",
    journal = "SciPost Phys.",
    volume = "18",
    number = "1",
    pages = "028",
    year = "2025"
}

@article{Kaidi:2025hyr,
    author = "Kaidi, Justin and Shi, Xiaoyi and Shimamori, Soichiro and Sun, Zhengdi",
    title = "{The SymTFT for $N$-ality defects: Part I}",
    eprint = "2509.19429",
    archivePrefix = "arXiv",
    primaryClass = "hep-th",
    month = "9",
    year = "2025"
}

@article{Lu:2024lzf,
    author = "Lu, Da-Chuan and Sun, Zhengdi and Zhang, Zipei",
    title = "{Exploring G-ality defects in 2-dim QFTs}",
    eprint = "2406.12151",
    archivePrefix = "arXiv",
    primaryClass = "hep-th",
    doi = "10.1007/JHEP11(2025)081",
    journal = "JHEP",
    volume = "11",
    pages = "081",
    year = "2025"
}

@article{Lu:2024ytl,
    author = "Lu, Da-Chuan and Sun, Zhengdi and You, Yi-Zhuang",
    title = "{Realizing triality and $p$-ality by lattice twisted gauging in (1+1)d quantum spin systems}",
    eprint = "2405.14939",
    archivePrefix = "arXiv",
    primaryClass = "cond-mat.str-el",
    doi = "10.21468/SciPostPhys.17.5.136",
    journal = "SciPost Phys.",
    volume = "17",
    number = "5",
    pages = "136",
    year = "2024"
}

@article{Maeda:2025rxc,
    author = "Maeda, Jun and Oishi, Tsubasa",
    title = "{$N$-ality symmetry and SPT phases in (1+1)d}",
    eprint = "2504.20151",
    archivePrefix = "arXiv",
    primaryClass = "hep-th",
    reportNumber = "YITP-25-35, KUNS-3038",
    month = "4",
    year = "2025"
}

@article{Ando:2024hun,
    author = "Ando, Takamasa",
    title = "{A journey on self-$G$-ality}",
    eprint = "2405.15648",
    archivePrefix = "arXiv",
    primaryClass = "cond-mat.str-el",
    month = "5",
    year = "2024"
}

@article{Lu:2025gpt,
    author = "Lu, Da-Chuan and Sun, Zhengdi and Zhang, Zipei",
    title = "{SymSETs and self-dualities under gauging non-invertible symmetries}",
    eprint = "2501.07787",
    archivePrefix = "arXiv",
    primaryClass = "hep-th",
    month = "1",
    year = "2025"
}

@article{Lu:2022ver,
    author = "Lu, Da-Chuan and Sun, Zhengdi",
    title = "{On triality defects in 2d CFT}",
    eprint = "2208.06077",
    archivePrefix = "arXiv",
    primaryClass = "hep-th",
    doi = "10.1007/JHEP02(2023)173",
    journal = "JHEP",
    volume = "02",
    pages = "173",
    year = "2023"
}

@article{Etingof:2009yvg,
    author = "Etingof, Pavel and Nikshych, Dmitri and Ostrik, Victor and Meir, with an appendix by Ehud",
    title = "{Fusion categories and homotopy theory}",
    eprint = "0909.3140",
    archivePrefix = "arXiv",
    primaryClass = "math.QA",
    month = "9",
    year = "2009"
}

@article{Gelaki:2009blp,
    author = "Gelaki, Shlomo and Naidu, Deepak and Nikshych, Dmitri",
    title = "{Centers of graded fusion categories}",
    eprint = "0905.3117",
    archivePrefix = "arXiv",
    primaryClass = "math.QA",
    month = "5",
    year = "2009"
}

@article{Thorngren:2021yso,
    author = "Thorngren, Ryan and Wang, Yifan",
    title = "{Fusion category symmetry. Part II. Categoriosities at c = 1 and beyond}",
    eprint = "2106.12577",
    archivePrefix = "arXiv",
    primaryClass = "hep-th",
    doi = "10.1007/JHEP07(2024)051",
    journal = "JHEP",
    volume = "07",
    pages = "051",
    year = "2024"
}

@article{Choi:2022jqy,
    author = "Choi, Yichul and Lam, Ho Tat and Shao, Shu-Heng",
    title = "{Noninvertible Global Symmetries in the Standard Model}",
    eprint = "2205.05086",
    archivePrefix = "arXiv",
    primaryClass = "hep-th",
    reportNumber = "YITP-SB-2022-21, MIT-CTP/5433",
    doi = "10.1103/PhysRevLett.129.161601",
    journal = "Phys. Rev. Lett.",
    volume = "129",
    number = "16",
    pages = "161601",
    year = "2022"
}

@article{Choi:2022zal,
    author = "Choi, Yichul and Cordova, Clay and Hsin, Po-Shen and Lam, Ho Tat and Shao, Shu-Heng",
    title = "{Non-invertible Condensation, Duality, and Triality Defects in 3+1 Dimensions}",
    eprint = "2204.09025",
    archivePrefix = "arXiv",
    primaryClass = "hep-th",
    reportNumber = "YITP-SB-2022-16, MIT/CTP-5423, YITP-SB-2022-16, MIT/CTP-5423",
    doi = "10.1007/s00220-023-04727-4",
    journal = "Commun. Math. Phys.",
    volume = "402",
    number = "1",
    pages = "489--542",
    year = "2023"
}

@article{Cordova:2022ieu,
    author = "Cordova, Clay and Ohmori, Kantaro",
    title = "{Noninvertible Chiral Symmetry and Exponential Hierarchies}",
    eprint = "2205.06243",
    archivePrefix = "arXiv",
    primaryClass = "hep-th",
    doi = "10.1103/PhysRevX.13.011034",
    journal = "Phys. Rev. X",
    volume = "13",
    number = "1",
    pages = "011034",
    year = "2023"
}

@article{Hayashi:2022fkw,
    author = "Hayashi, Yui and Tanizaki, Yuya",
    title = "{Non-invertible self-duality defects of Cardy-Rabinovici model and mixed gravitational anomaly}",
    eprint = "2204.07440",
    archivePrefix = "arXiv",
    primaryClass = "hep-th",
    reportNumber = "YITP-22-12",
    doi = "10.1007/JHEP08(2022)036",
    journal = "JHEP",
    volume = "08",
    pages = "036",
    year = "2022"
}

@article{Kaidi:2021xfk,
    author = "Kaidi, Justin and Ohmori, Kantaro and Zheng, Yunqin",
    title = "{Kramers-Wannier-like Duality Defects in (3+1)D Gauge Theories}",
    eprint = "2111.01141",
    archivePrefix = "arXiv",
    primaryClass = "hep-th",
    doi = "10.1103/PhysRevLett.128.111601",
    journal = "Phys. Rev. Lett.",
    volume = "128",
    number = "11",
    pages = "111601",
    year = "2022"
}

@article{Kaidi:2023maf,
    author = "Kaidi, Justin and Nardoni, Emily and Zafrir, Gabi and Zheng, Yunqin",
    title = "{Symmetry TFTs and anomalies of non-invertible symmetries}",
    eprint = "2301.07112",
    archivePrefix = "arXiv",
    primaryClass = "hep-th",
    doi = "10.1007/JHEP10(2023)053",
    journal = "JHEP",
    volume = "10",
    pages = "053",
    year = "2023"
}

@article{Tachikawa:2017gyf,
    author = "Tachikawa, Yuji",
    title = "{On gauging finite subgroups}",
    eprint = "1712.09542",
    archivePrefix = "arXiv",
    primaryClass = "hep-th",
    reportNumber = "IPMU-17-0183",
    doi = "10.21468/SciPostPhys.8.1.015",
    journal = "SciPost Phys.",
    volume = "8",
    number = "1",
    pages = "015",
    year = "2020"
}

@article{Hsin:2018vcg,
    author = "Hsin, Po-Shen and Lam, Ho Tat and Seiberg, Nathan",
    title = "{Comments on One-Form Global Symmetries and Their Gauging in 3d and 4d}",
    eprint = "1812.04716",
    archivePrefix = "arXiv",
    primaryClass = "hep-th",
    doi = "10.21468/SciPostPhys.6.3.039",
    journal = "SciPost Phys.",
    volume = "6",
    number = "3",
    pages = "039",
    year = "2019"
}

@article{Thorngren:2016hdm,
    author = "Thorngren, Ryan and Else, Dominic V.",
    title = "{Gauging Spatial Symmetries and the Classification of Topological Crystalline Phases}",
    eprint = "1612.00846",
    archivePrefix = "arXiv",
    primaryClass = "cond-mat.str-el",
    doi = "10.1103/PhysRevX.8.011040",
    journal = "Phys. Rev. X",
    volume = "8",
    number = "1",
    pages = "011040",
    year = "2018"
}

@article{Else:2018eas,
    author = "Else, Dominic V. and Thorngren, Ryan",
    title = "{Crystalline topological phases as defect networks}",
    eprint = "1810.10539",
    archivePrefix = "arXiv",
    primaryClass = "cond-mat.str-el",
    doi = "10.1103/PhysRevB.99.115116",
    journal = "Phys. Rev. B",
    volume = "99",
    number = "11",
    pages = "115116",
    year = "2019"
}

@article{Oshikawa:1992smv,
    author = "Oshikawa, M.",
    title = "{Hidden Z2*Z2 symmetry in quantum spin chains with arbitrary integer spin}",
    doi = "10.1088/0953-8984/4/36/019",
    journal = "J. Phys. Condens. Matter",
    volume = "4",
    number = "36",
    pages = "7469",
    year = "1992"
}

@article{Li:2023ani,
    author = "Li, Linhao and Oshikawa, Masaki and Zheng, Yunqin",
    title = "{Noninvertible duality transformation between symmetry-protected topological and spontaneous symmetry breaking phases}",
    eprint = "2301.07899",
    archivePrefix = "arXiv",
    primaryClass = "cond-mat.str-el",
    doi = "10.1103/PhysRevB.108.214429",
    journal = "Phys. Rev. B",
    volume = "108",
    number = "21",
    pages = "214429",
    year = "2023"
}

@article{Kennedy:1992ifl,
    author = "Kennedy, Tom and Tasaki, Hal",
    title = "{Hidden Z2{\texttimes}Z2 symmetry breaking in Haldane-gap antiferromagnets}",
    doi = "10.1103/PhysRevB.45.304",
    journal = "Phys. Rev. B",
    volume = "45",
    number = "1",
    pages = "304",
    year = "1992"
}

@article{Kennedy:1992tke,
    author = "Kennedy, Tom and Tasaki, Hal",
    title = "{Hidden symmetry breaking and the Haldane phase inS=1 quantum spin chains}",
    doi = "10.1007/bf02097239",
    journal = "Commun. Math. Phys.",
    volume = "147",
    number = "3",
    pages = "431--484",
    year = "1992"
}

@article{Aksoy:2023hve,
    author = {Aksoy, {\"O}mer Mert and Mudry, Christopher and Furusaki, Akira and Tiwari, Apoorv},
    title = "{Lieb-Schultz-Mattis anomalies and web of dualities induced by gauging in quantum spin chains}",
    eprint = "2308.00743",
    archivePrefix = "arXiv",
    primaryClass = "cond-mat.str-el",
    doi = "10.21468/SciPostPhys.16.1.022",
    journal = "SciPost Phys.",
    volume = "16",
    number = "1",
    pages = "022",
    year = "2024"
}

@article{Alavirad:2019iea,
    author = "Alavirad, Yahya and Barkeshli, Maissam",
    title = "{Anomalies and unusual stability of multicomponent Luttinger liquids in Zn{\texttimes}Zn spin chains}",
    eprint = "1910.00589",
    archivePrefix = "arXiv",
    primaryClass = "cond-mat.str-el",
    doi = "10.1103/PhysRevB.104.045151",
    journal = "Phys. Rev. B",
    volume = "104",
    number = "4",
    pages = "045151",
    year = "2021"
}

@article{Cao:2024qjj,
    author = "Cao, Weiguang and Li, Linhao and Yamazaki, Masahito",
    title = "{Generating lattice non-invertible symmetries}",
    eprint = "2406.05454",
    archivePrefix = "arXiv",
    primaryClass = "cond-mat.str-el",
    doi = "10.21468/SciPostPhys.17.4.104",
    journal = "SciPost Phys.",
    volume = "17",
    number = "4",
    pages = "104",
    year = "2024"
}

@article{Choi:2024rjm,
    author = "Choi, Yichul and Sanghavi, Yaman and Shao, Shu-Heng and Zheng, Yunqin",
    title = "{Non-invertible and higher-form symmetries in 2+1d lattice gauge theories}",
    eprint = "2405.13105",
    archivePrefix = "arXiv",
    primaryClass = "cond-mat.str-el",
    doi = "10.21468/SciPostPhys.18.1.008",
    journal = "SciPost Phys.",
    volume = "18",
    number = "1",
    pages = "008",
    year = "2025"
}

@article{Roumpedakis:2022aik,
    author = "Roumpedakis, Konstantinos and Seifnashri, Sahand and Shao, Shu-Heng",
    title = "{Higher Gauging and Non-invertible Condensation Defects}",
    eprint = "2204.02407",
    archivePrefix = "arXiv",
    primaryClass = "hep-th",
    reportNumber = "YITP-SB-2022-14",
    doi = "10.1007/s00220-023-04706-9",
    journal = "Commun. Math. Phys.",
    volume = "401",
    number = "3",
    pages = "3043--3107",
    year = "2023"
}

@article{Seifnashri:2024dsd,
    author = "Seifnashri, Sahand and Shao, Shu-Heng",
    title = "{Cluster State as a Noninvertible Symmetry-Protected Topological Phase}",
    eprint = "2404.01369",
    archivePrefix = "arXiv",
    primaryClass = "cond-mat.str-el",
    reportNumber = "YITP-SB-2024-05",
    doi = "10.1103/PhysRevLett.133.116601",
    journal = "Phys. Rev. Lett.",
    volume = "133",
    number = "11",
    pages = "116601",
    year = "2024"
}

@article{Gorantla:2024ocs,
    author = "Gorantla, Pranay and Shao, Shu-Heng and Tantivasadakarn, Nathanan",
    title = "{Tensor Networks for Noninvertible Symmetries in 3+1D and Beyond}",
    eprint = "2406.12978",
    archivePrefix = "arXiv",
    primaryClass = "quant-ph",
    reportNumber = "YITP-SB-2024-11",
    doi = "10.1103/p32z-v884",
    journal = "Phys. Rev. X",
    volume = "15",
    number = "4",
    pages = "041006",
    year = "2025"
}

@article{Yao:2025iia,
    author = "Yao, Ching-Yu",
    title = "{Lattice Translation Modulated Symmetries and TFTs}",
    eprint = "2510.03889",
    archivePrefix = "arXiv",
    primaryClass = "cond-mat.str-el",
    month = "10",
    year = "2025"
}

@article{Cao:2023rrb,
    author = "Cao, Weiguang and Jia, Qiang",
    title = "{Symmetry TFT for subsystem symmetry}",
    eprint = "2310.01474",
    archivePrefix = "arXiv",
    primaryClass = "hep-th",
    reportNumber = "KIAS-P23043",
    doi = "10.1007/JHEP05(2024)225",
    journal = "JHEP",
    volume = "05",
    pages = "225",
    year = "2024"
}

@article{Cao:2025qhg,
    author = "Cao, Weiguang and Yamazaki, Masahito and Li, Linhao",
    title = "{Duality viewpoint of noninvertible symmetry protected topological phases}",
    eprint = "2502.20435",
    archivePrefix = "arXiv",
    primaryClass = "cond-mat.str-el",
    month = "2",
    year = "2025"
}
\bibliographystyle{utphys}
\end{document}